\documentclass[prd,aps,amsfonts,eqsecnum,nofootinbib,notitlepage,longbibliography,10pt]{revtex4-1} 

\usepackage{graphicx}
\usepackage{tabularx}
\usepackage{braket}
\usepackage{amssymb}
\usepackage{amsmath}
\usepackage{bm}
\usepackage{upgreek}
\usepackage[driverfallback=dvipdfm]{hyperref}
\hypersetup{colorlinks=true,breaklinks,urlcolor=blue,linkcolor=blue,citecolor=blue}

\usepackage{chngcntr}
\counterwithout{equation}{section}

\usepackage{pdfpages}
\usepackage{tikz}
\makeatletter
\AtBeginDocument{\let\LS@rot\@undefined}
\makeatother

\newcommand{\revise}[1]{#1}

\newcommand{\comment}[1]{}
\usepackage{xcolor}
\definecolor{ginger}{rgb}{0.69, 0.4, 0.0}

\newcommand{\lr}[1]{\left( #1\right)}

\newcommand{\alr}[1]{\left\langle #1\right\rangle}
\newcommand{\norm}[1]{\left\lVert#1\right\rVert}
\newcommand{\abs}[1]{\left\lvert#1\right\rvert}

\newcommand{\ii}{\mathrm{i}}
\newcommand{\ee}{\mathrm{e}}
\newcommand{\dd}{\mathrm{d}}

\newcommand{\where}{\quad {\rm where}\quad}
\newcommand{\order}{\mathrm{O}}

\newcommand{\poly}[1]{\mathrm{poly}\lr{#1} }
\newcommand{\polylog}{\mathrm{polylog} }
\newcommand{\fkd}{\mathfrak{d}}

\newcommand{\OO}{\mathcal{O}}

\newcommand{\kpsi}{\ket{\psi}}
\newcommand{\kzero}{\ket{\bm{0}}}

\makeatletter
\renewcommand{\p@subsection}{}
\renewcommand{\p@subsubsection}{}
\makeatother

\usepackage{amsthm}
\newtheorem{thm}{Theorem}
\newtheorem{cor}[thm]{Corollary}

\begin{document}

\title{Fast quantum computation with all-to-all Hamiltonians}
\author{Chao Yin}\email{chaoyin@stanford.edu}
\affiliation{Department of Physics, Stanford University, Stanford, California 94305, USA}

\date{\today}

\begin{abstract}
All-to-all interactions arise naturally in many areas of theoretical physics and across diverse experimental quantum platforms, motivating a systematic study of their information-processing power. Assuming each pair of qubits interacts with $\mathrm{O}(1)$ strength, \revise{programmable} time-dependent all-to-all Hamiltonians can simulate arbitrary all-to-all quantum circuits, performing quantum computation in time proportional to the circuit depth. We show that this naive correspondence is far from optimal: all-to-all Hamiltonians can process information on much shorter timescales.

First, we prove that any two-qubit gate can be simulated by all-to-all Hamiltonians on $N$ qubits in time $\mathrm{O}(1/N)$ (up to factor $N^{\delta}$ with an arbitrarily small constant $\delta>0$), with polynomially small error $1/\mathrm{poly}(N)$. Immediate consequences include: 
1) Certain $\mathrm{O}(N)$-qubit unitaries and entangled states, such as the multiply-controlled Toffoli gate and the GHZ and W states, can be generated in $\mathrm{O}(1/N)$ time; 
2) Information could propagate in a fast way that saturates known Lieb-Robinson bounds in strongly power-law interacting systems. 

Our second main result proves that any depth-$D$ quantum circuit can be simulated by a randomized Hamiltonian protocol in time $T=\mathrm{O}(D/\sqrt{N})$, with constant space overhead and polynomially small error. \revise{Applied to circuit ensembles forming unitary designs and pseudorandom unitaries, this simulation gives an operational proof of the fast scrambling conjecture for dense Hamiltonians.} 

The techniques underlying our results depart fundamentally from the existing literature on parallelizing commuting gates: We rely crucially on non-commuting Hamiltonians and draw on diverse physical ideas.
\end{abstract}

\maketitle

\section{Introduction}

Promising advantages on difficult problems \cite{Shor99}, quantum computing has witnessed tremendous developments in the current generation. The canonical model of quantum computation is quantum circuits \cite{Nielsen_Chuang_2010}, where any pair of qubits can interact by a two-qubit gate assuming all-to-all connectivity. In reality, however, the qubits are arranged in our 3-dimensional world, so gates acting on faraway qubits require extra cost. For example, one can move the qubits so that they can interact with new neighbors \cite{Rydberg_processor23,ion_racetrack23}, or use SWAP gates to route quantum information directly \cite{routing_10,routing_overhead20,routing_Maryland23}. These routing approaches induce an overhead that scales with the system size. Another approach, however, is to couple all qubits to some global modes of photons or phonons, so that an effective \emph{all-to-all Hamiltonian} for the qubits is induced. This situation
arises naturally in quantum devices involving optical cavities \cite{cavity_metro_scram23,cavity_graphstate24}, trapped ions \cite{ion_longrange12,ion_rev21}, and superconducting qubits \cite{SC_all2all20,all2all_supercond}. 

Historically, all-to-all Hamiltonians also play an important role in diverse areas of physics, ranging from quantum optics \cite{Dickemodel54}, nuclear physics \cite{LMG65}, spin glass \cite{SK75,SK_Parisi79} to quantum gravity \cite{SYK93,SYK_talk15,SYK_remark16}. The mean-field nature of all-to-all Hamiltonians makes them more tractable analytically, paving the way to understanding more complicated situations with extra spatial structure. Moreover, understanding all-to-all interactions provides insight into more general long-range Hamiltonians with slowly decaying couplings \cite{longrange_rev23}. Due to such importance of all-to-all Hamiltonians, it is demanding to understand their \revise{dynamics from a computational-complexity viewpoint}.

With a time-dependent all-to-all Hamiltonian $H=\sum_{ij}H_{ij}$, one can realize information processing quantum circuits by turning on the coupling term $H_{ij}$ for some time for a two-qubit gate on $i,j$. 
However, this approach simulating circuits by Hamiltonians does not exploit the full power of all-to-all interactions: An $N$-qubit Hamiltonian $H$ contains $\mathrm{\Theta}(N^2)$ interactions, while only $\order(N)$ of them are turned on at a given time, because each qubit is involved in at most one gate a time in the circuit model. This raises the following question: Given a quantum information processing task, say a unitary $U$, what is the shortest evolution time $T$ for an all-to-all Hamiltonian to generate $U$? If $U$ can be realized by a depth-$D$ circuit, then a $T=\mathrm{\Theta}(D)$ Hamiltonian evolution would suffice assuming $\order(1)$ interaction strength per qubit pair, but can Hamiltonians be asymptotically faster $T\ll D$? Such a speedup is particularly demanding for near-term quantum devices to finish useful computation before the decoherence timescale.

This question has been studied in the context of quantum circuits aided by global entangling gates \cite{fanout05,fanout16,globalgate_18,globalgate_21,globalgate_lognClif22,Clifford_constcost22,global_gate23,global_gate_memory24,global_gate25}, where some commuting gates 
are allowed to be executed in parallel, corresponding to an all-to-all commuting Hamiltonian. The seminal work \cite{Clifford_constcost22} shows that $T=\order(1)$ Hamiltonian evolution can generate any Clifford circuit and any multiply-controlled Toffoli gate, both of which require a circuit depth (without global gates) that grows with the system size $N$. However, Clifford circuits are classically simulable \cite{gottesman_knill98}, so it is unclear how the speedup applies to interesting quantum tasks. The speedup for the multiply-controlled Toffoli is also a mild logarithmic factor $\order(\log N)$ comparing to circuits with $\mathrm{\Theta}(N)$ ancillae (i.e., constant space overhead) \cite{Nielsen_Chuang_2010}. 

In this work, we show that all-to-all Hamiltonians can be \emph{polynomially} faster \revise{---achieving $T\le D N^{-\Omega(1)}$---} for a wide range of quantum information processing tasks, which significantly improve upon previous works. We rigorously prove two main results. First, interesting families of circuits including the multiply-controlled Toffoli can actually be realized in time $T\sim 1/N$\footnote{Here $\sim$ hides factors like $N^\delta$ with an arbitrarily small constant $\delta$.} with constant space overhead and a polynomially small error $1/\mathrm{poly}(N)$, around $N$ times faster than previous results \cite{Clifford_constcost22}. Although $T$ vanishes with $N$, we want it to be as small as possible because a complicated task may contain many subroutines of our protocol each taking that amount of time\revise{; see the further justifications below}. The $1/\mathrm{poly}(N)$ error also ensures a small total error as long as the total task has polynomial size. 
Our $T\sim 1/N$ protocol has many implications that we will discuss.
\comment{
Many-body entangled states like the Greenberger-Horne-Zeilinger (GHZ) states \cite{GHZ89} and W states \cite{W_state} can also be prepared in the same timescale, providing at least a $\sqrt{N}$ speedup than known results \cite{Wprotocol_gorshkov20}. We also show that for simulating any quantum circuit one can always trade larger space (i.e., more ancilla qubits) for shorter time that preserves the total spacetime volume.
}
Although this $N$-factor speedup holds for specific problems, our second main result shows that Hamiltonians could be $\sqrt{N}$ times faster for general purpose information-processing tasks: \emph{Any} quantum circuit of polynomial depth $D$ can be simulated by a randomized Hamiltonian evolution in time $T\sim D/\sqrt{N}$ with constant space overhead, and average simulation error $1/\mathrm{poly}(N)$. \revise{As a corollary, we provide an operational proof of the fast scrambling conjecture \cite{fast_scramblers08,fast_scrambler13,OTOC_chaosbound16,randomU_25} in the family of dense Hamiltonians like the Sachdev-Ye-Kitaev (SYK) model \cite{SYK93,SYK_talk15,SYK_remark16}: an ``approximate'' Haar ensemble can be generated in ``almost'' logarithmic time when the Hamiltonian is extensively normalized, $\norm{H}=\order(N)$.}

The ideas behind our results appear fundamentally different from previous works on global entangling gates that aim to parallelize commuting gates \cite{Clifford_constcost22}. Our approach crucially involves \emph{non-commuting} Hamiltonians, and borrows classic ideas in physics ranging from semiclassical quantum mechanics, (spin) squeezing \cite{squeeze93,squeeze_rev11,q_optics_book}, Mølmer-Sørensen gates \cite{MS99,MS99_}, to spin waves. Note that some of our ideas have appeared in the author's previous work \cite{myGHZ} on a heuristic protocol, which we generalize here for a rigorous treatment. Our first main result also settles the information speed limit problem for power-law interactions by matching an existing Lieb-Robinson bound \cite{Wprotocol_gorshkov20}. Conversely, this implies that our $T\sim 1/N$ protocol is asymptotically optimal and cannot be improved. Since Lieb-Robinson bound \cite{Lieb1972} plays a fundamental role in analyzing many-body systems based on locality \cite{ourreview}, our results shed light on understanding power-law systems in general, such as their simulation complexity \cite{HHKL,power_simu19,power_simu23}, entanglement structure \cite{area_law07,area_1dlong20} and thermalization timescales \cite{our_metastable25}.

\section{Preliminaries}

Here we set up the problem precisely.
We consider $K$-local all-to-all Hamiltonians on $N_{\rm tot}$ qubits rather than just $2$-local ones: \begin{equation}\label{eq:Hint=}
H_{\rm int}(t)=\sum_{k=2}^K N_{\rm tot}^{2-k} \sum_{ i_1<i_2<\cdots<i_k} \sum_{a_1\cdots a_k}J_{i_1\cdots i_k}^{a_1\cdots a_k}(t) X_{i_1}^{a_1}\cdots X_{i_k}^{a_k}, \where \sum_{a_1\cdots a_k}\abs{J_{i_1\cdots i_k}^{a_1\cdots a_k}(t)}\le 1.
\end{equation} 
Here $X^a_i$ ($a=1,2,3$) are the Pauli matrices $X_i,Y_i,Z_i$ for qubit $i\in \mathbb{Z}$. We assume the coefficients $J_{i_1\cdots i_k}^{a_1\cdots a_k}(t)$ can depend arbitrarily in time as long as \eqref{eq:Hint=} is obeyed. The system is evolved under $H(t)=H_{\rm int}(t)+H_{\rm rot}(t)$ that also includes instantaneous single-qubit rotations $H_{\rm rot}(t)=\sum_i\sum_a h_i^a(t)X_i^a$.
We make \revise{several} remarks: \begin{itemize}
    \item \revise{We first discuss the global normalization of $H$ in the $K=2$ case. The convention in \eqref{eq:Hint=} is chosen so that the naive Hamiltonian simulation of a depth-$D$ circuit costs $T=\Theta(D)$, giving a direct comparison with the circuit model, even though $\norm{H_{\rm int}}=\order(N_{\rm tot}^2)$ can be superextensive. The problem of finding the fastest protocols within the Hamiltonian family \eqref{eq:Hint=} is independent of this convention, since a global rescaling of $H$ simply rescales time. For example, the physical Hamiltonian $H_{\rm phys}=N_{\rm tot}^{-1}H$ for trapped ions in the all-to-all limit \cite{ion_rev21} has a different normalization: our $T\sim 1/N_{\rm tot}$ protocol below becomes $T_{\rm phys}\sim 1$, while the naive simulation of circuits becomes $T_{\rm phys}\sim N_{\rm tot}$. We will also see that our $\sqrt{N}$-speedup protocol matches the fastest rescaled timescale suggested by the fast scrambling conjecture in Hamiltonian family \eqref{eq:Hint=}.

    \item Although the superextensive value of $\norm{H_{\rm int}}$ already suggests from energy-time considerations that naive circuit simulation is suboptimal, all-to-all Hamiltonians are usually treated as analog evolutions rather than as a digital-computation resource. Our contribution is to show that programmable coupling coefficients enable fast compilation of general unitaries.
    
    \item If the comparison class instead fixes the global norm $\norm{H_{\rm int}}$, rather than the pairwise coupling strength in \eqref{eq:Hint=}, then our protocols do not outperform circuit protocols under that single resource measure: one could turn on fewer couplings at a time but make their individual coefficients larger. However, such all-to-all circuit protocols often require routing overheads that are not captured by the single resource count $\norm{H_{\rm int}}$: it is not straightforward to realize a Hamiltonian with expander-graph-like connectivity. As an example, for trapped ions with tunable interaction range \cite{ion_rev21}, a more natural comparison is between the all-to-all limit $H_{\rm phys}=N_{\rm tot}^{-1}H$ and geometrically local Hamiltonians with $\order(1)$ local strength. In this comparison, we will see that our $T_{\rm phys}\sim 1$ protocol for generating globally entangled states is still faster, because merely propagating information across a geometrically local system takes at least polynomial time.} 
    
    \item \revise{In the general model \eqref{eq:Hint=}, we allow multibody interactions, which can be realized experimentally \cite{ion_34body,ion_Nbody22,ion_Nbody23,clock_multibody18,cavity_34body,GHZlike24,constraint_Nbody25}.} The factor $N_{\rm tot}^{2-k}$ in \eqref{eq:Hint=} ensures that the total Hamiltonian norm acting nontrivially on a qubit pair is $\order(1)$. \revise{This normalization} is fairly stringent: not just the total norm $\norm{H_k}$ for $k$-body couplings should be comparable, but each individual coupling is polynomially small. Note that our $\sqrt{N}$-speedup result only requires $K=2$.
    
    \item We assume arbitrarily strong on-site fields $h_i^a$ following \cite{Clifford_constcost22,myGHZ}, because their effect can be absorbed by $H_{\rm int}$ in an interaction picture. This assumption is also reasonable in reality, where single-qubit rotations commonly operate on timescales far shorter than than entangling operations \cite{Rydberg_rev20,ion_rev21,cavity_Zeyang22,cavity_BCS24}.
\end{itemize}

We aim to use $H(t)$ to simulate quantum circuits. We say unitary $U$ is a depth-$D$ quantum circuit if $U=U_D\cdots U_2 U_1$ where each layer $U_m$ consists of nonoverlapping two-qubit controlled-Z (CZ) gates $\mathrm{CZ}=I-2\ket{11}\bra{11}$, followed by arbitrary single-qubit rotations. Since they form a universal gate set, our results generalize to circuits with general finite-range entangling gates.

In our protocols, we use $N=\mathrm{\Theta}(N_{\rm tot})$ qubits, a finite portion of the system, as ancillae that start the evolution in the all-zero state $\kzero$, and approximately returns to the same state. Information is processed in the other $N_{\rm d}=N_{\rm tot}-N$ data qubits. \revise{Most of our results below use} a constant space overhead $N_{\rm d}=\mathrm{\Theta}(N)$. \revise{Since our focus is the fundamental speed limit of all-to-all Hamiltonian dynamics, we assume perfect noiseless evolution throughout.}

\section{$1/N$-time protocols}

We are ready to state our first main result:
\begin{thm}\label{thm:CZ}
    For any constants $\delta_{\rm T}\in(0,1)$ and locality $K\ge 2$, the following holds for sufficiently large $N$.
    For a set of data qubits $\{0\}\cup S$,
    there exists a Hamiltonian protocol $H(t)$ \revise{with normalization \eqref{eq:Hint=}} that simulates a bunch of CZ gates $\prod_i\mathrm{CZ}_{0,i\in S}$ with polynomially small error \begin{equation}\label{eq:eps=N}
        \norm{\lr{\mathcal{T}\ee^{-\ii\int^T_0 \dd tH(t)}-\prod_i\mathrm{CZ}_{0,i\in S}}\kpsi\otimes\kzero} \le N^{2-\delta_{\rm T}(\sqrt{K}-1)/2},
    \end{equation} 
    for any $\ket{\psi}$ on data qubits, in total evolution time \begin{equation}
        T\le N^{-1+\delta_{\rm T}}.
    \end{equation} Here $\mathcal{T}$ is time-ordering.
\end{thm}

We illustrate the idea shortly, and refer to Theorem 3.1 and Corollary 5.2 in Supplemental Material (SM) \cite{SM} for the formal proof.
\eqref{eq:eps=N} requires a sufficiently large $K$ independent of $N$ to get a desired polynomially small error. Nevertheless, the ideas of Theorem \ref{thm:CZ} could apply more generally. Indeed, the single CZ gate version of Theorem \ref{thm:CZ} is a generalization of the
$2$-local Hamiltonian protocol in \cite{myGHZ}, which works well in practice and yields a $\sim 15$-times speedup for thousands of qubits. The generalization to $K$-locality here is primarily aiming for a rigorous error bound\revise{, and we leave open whether provably fast protocols can already be established for the $2$-local case.} One strategy is to \revise{try to} simulate a $K$-local Hamiltonian by $2$-local ones using Floquet engineering \cite{GHZlike24} \revise{whose error could be bounded \cite{magnus_comm25}}. 
For example, $\ee^{-\ii T X^2}\ee^{-\ii T Y^2}$ induces an effective Hamiltonian $H_{\rm eff}=-\ii T [X^2,Y^2]+\cdots =4T XYZ + \cdots$ containing $\ge 3$-body terms. Here $X=\sum_i X_i$ and $Y,Z$ defined similarly. With such multi-body interactions, Theorem \ref{thm:CZ} suggests that a $4$-local protocol for example may be much more accurate than $2$-local ones like \cite{myGHZ}, while remaining experimentally feasible. 

\subsection{Applications}

Theorem \ref{thm:CZ} shows that the unbounded fan-out gate \cite{fanout05}, equivalent to $\prod_i\mathrm{CZ}_{0,i\in S}$, 
can be simulated in time $T\sim 1/N$. This immediately implies a $T\sim 1/N$ protocol that prepares the Greenberger-Horne-Zeilinger (GHZ) state $\frac{1}{\sqrt{2}}(\ket{0\cdots0}+\ket{1\cdots 1})$, $\sim N$ times faster than previously known rigorous protocols using all-to-all interactions \cite{ourreview,myGHZ}. 
More generally, Theorem \ref{thm:CZ} implies fast protocols for the following families of circuits and states (see more general statements in Section 5.2,5.3 of SM \cite{SM}). 

\begin{cor}\label{cor:GHZ_W}
    The multiply-controlled Toffoli on $n+1\le N_{\rm d}$ data qubits $\ket{z}\ket{y}\rightarrow \ket{z}\ket{y\oplus g(z)} $ where $g(z)=1$ iff $z_1=\cdots=z_n=1$, can be simulated with $1/\mathrm{poly}(N)$ error by a Hamiltonian protocol $H(t)$ in time $T\sim 1/N$. The GHZ state and W state $\frac{1}{\sqrt{n}}(\ket{10\cdots0}+\ket{010\cdots0}+\cdots +\ket{0\cdots01})$ can be prepared using the same amount of resources.
\end{cor}

In particular, the W-state protocol is $\sim \sqrt{N}$ times faster than the previously known ones \cite{Wprotocol_gorshkov20}, while the multiply-controlled Toffoli protocol is $\sim N$ times faster than \cite{Clifford_constcost22}. The idea is that $\mathrm{\Theta}(\log N)$ unbounded fan-out gates are able to extract the Hamming weight information $|z|$ of bitstrings $z$ \cite{fanout05,fanout16,Clifford_constcost22}, while the multiply-controlled Toffoli and W state are effectively extracting a particular Hamming weight.

\comment{
If we use much more ancilae than data qubits $N\gg N_{\rm d}$, Theorem \ref{thm:CZ} further provides a general strategy to trade space for time using Hamiltonians, by sequentially simulating each gate of a target unitary (see Corollary 5.1 in SM \cite{SM}):
\begin{cor}\label{cor:space_time}
    Any depth-$D$ circuit on $N_{\rm d}\ll N$ qubits can be simulated by a $T\sim D N_{\rm d}/N\ll D$ Hamiltonian protocol with $1/\poly{N}$ error.
\end{cor}
Intriguingly, one can always make the time $T$ vanish by choosing a sufficiently large space overhead.
}

\revise{A second application concerns fundamental information-speed limits.}
For local Hamiltonians in constant spatial dimension $\mathsf{d}$, it is well-known from Lieb-Robinson bound \cite{Lieb1972} $\norm{\OO_i(T),\OO_j}\le \ee^{-\mu(r_{ij}-vT)}$ that information propagates with a bounded velocity $v$, where $\OO(T)$ is the Heisenberg evolution of operator $\OO$ and $r_{ij}$ is the distance between $i,j$. 
Such Lieb-Robinson bounds have been generalized to power-law interacting systems $H_{ij}\sim r_{ij}^{-\alpha}$ \cite{power_chen19,power_KSLRB20,power_hierarchy,power_LRB21} , which are shown to be tight for weakly long-range interactions $\alpha\ge \mathsf{d}$ based on fast protocols that match the bound 
\cite{power_hierarchy,power_GHZ21,power_yifan21}. For strongly long-range interactions $\alpha<\mathsf{d}$, however, it is unclear whether the bound derived in \cite{Wprotocol_gorshkov20} is tight, which claims \begin{equation}\label{eq:LRB}
    T=\Omega\lr{ N^{\frac{\alpha}{\mathsf{d}}-1}}
\end{equation} 
for information to propagate. See Section 5.4 in SM \cite{SM} that proves \eqref{eq:LRB} beyond $2$-local Hamiltonians \cite{Wprotocol_gorshkov20}. Theorem \ref{thm:CZ} implies that this bound is indeed tight:
\begin{cor}\label{cor:LRB}
    For any $\alpha<\mathsf{d}$ and any pair of qubits $i,j$, there exists a Hamiltonian $H(t)$ satisfying $\norm{H_{i'j'}(t)}\le r_{i'j'}^{-\alpha}\; (\forall i',j')$ where $H_{i'j'}$ is the part of $H$ that acts on $i',j'$, such that $T\sim N^{\frac{\alpha}{\mathsf{d}}-1}$ is sufficient to grow operators: \begin{equation}
        \norm{[X_i(T),X_j]}\ge 1.
    \end{equation}
\end{cor}

This is because one can get an all-to-all Hamiltonian by setting all couplings to the weakest strength, e.g. $H_{ij}\sim L^{-\alpha}$ for $2$-local Hamiltonians where $L=N^{1/\mathsf{d}}$ is the linear system size. The protocol of Corollary \ref{cor:LRB} is just Theorem \ref{thm:CZ} under this normalization, which establishes the optimality of both our protocol and Lieb-Robinson bound \eqref{eq:LRB}. 

\revise{SM \cite{SM} discusses further applications of Theorem \ref{thm:CZ}, including a general strategy for trading space for time and ancilla-assisted fast operator growth.}


\subsection{Proof sketch}
\begin{figure}[htbp]
    \centering
    \includegraphics[width=0.98\linewidth]{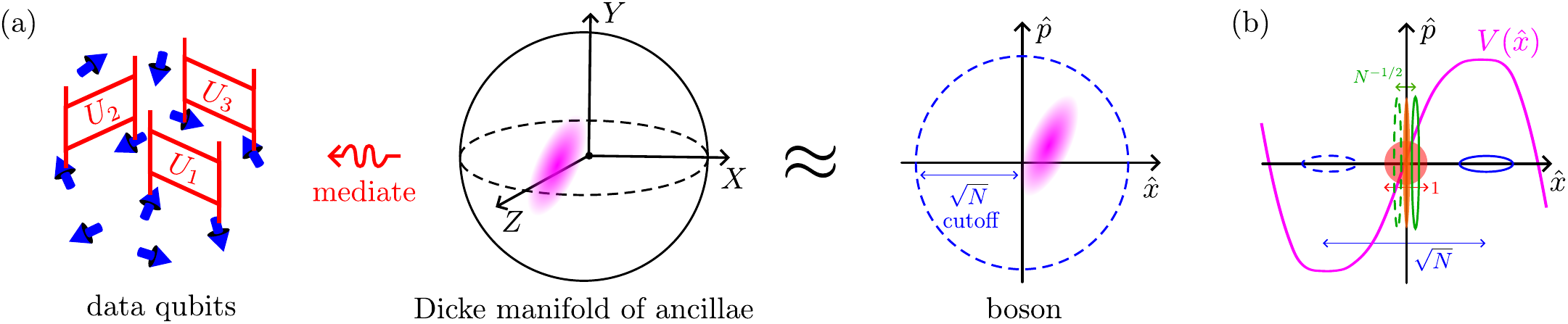}
    \caption{Sketch of the idea for Theorem \ref{thm:CZ}. (a) We mediate data-qubit interactions by $N$ ancila qubits, which simulates a boson mode with boson number $b^\dagger b\lesssim N$. (b) The protocol first amplifies the signal $Z_0$ to boson position $\hat{x}$ (red $\rightarrow$ yellow $\rightarrow$ green $\rightarrow$ blue, where solid/dashed contours correspond to $Z_0=\pm 1$), then engineers a potential $\pm V(\hat{x})$ controlled by the target data qubits $S$, and finally reverses the signal amplification step. }
    \label{fig:dicke_boson}
\end{figure}

Our main idea of Theorem \ref{thm:CZ} is to view the ancilla qubits as a boson mode and utilize bosonic squeezed states (see e.g. \cite{q_optics_book}). The $N$ ancillae work in their permutation-symmetric subspace of dimension $N+1$ known as the Dicke manifold, which is a large semiclassical spin with a spherical phase space. Close to its north pole, the large spin resembles a boson mode illustrated in Fig.~\ref{fig:dicke_boson}(a). The boson annihilation operator is obtained from the Holstein-Primakoff transformation \cite{HP_transf40} \begin{equation}\label{eq:HP}
    b = \frac{1}{2\sqrt{N}} \lr{1-\frac{N-Z}{2N}}^{-1/2} (X+\ii Y) = \frac{1}{2\sqrt{N}} \left[1+ \frac{N-Z}{4N}+ \frac{3}{8}\lr{\frac{N-Z}{2N}}^2+\cdots\right](X+\ii Y),
\end{equation} 
where 
polarization $X=\sum_{\text{ancilla }i} X_i$ ($Y$) of the ancillae qubits corresponds to position $\hat{x}=(b+b^\dagger)/\sqrt{2}$ (momentum $\hat{p}=(b-b^\dagger)/\sqrt{2}\ii$) of the boson with $[\hat{x},\hat{p}]=\ii$, and $N-Z=2b^\dagger b$ is the boson number. 

If we indeed use a boson mode as ancilla, the protocol is sketched in Fig.~\ref{fig:dicke_boson}(b) and works as follows. Starting from $\kzero$, i.e. no bosons, the protocol first squeezes the boson mode by Hamiltonian $H^{\rm b}_{\rm S}=\ii \frac{N}{2}(b^2-b^{\dagger 2})$, and then displaces it using $H^{\rm b}_{\rm CD}=\sqrt{2N}Z_0\otimes \hat{p}$ controlled by data qubit $0$. Because $H^{\rm b}_{\rm S}$ exponentially squeezes the state $\alr{\mathrm{\Delta} \hat{x}^2}\sim \ee^{-2NT_{\rm S}}$, a time $T_{\rm S}\sim 1/N$ suffices to yield $\alr{\mathrm{\Delta} \hat{x}^2}\sim 1/N$ so that a similar amount of time $T_{\rm CD}\sim 1/N$ in $H^{\rm b}_{\rm CD}$ can fully separate the wavepacket for the two values $Z_0=\pm 1$. We then time-reverse the first squeezing stage to pull away the two wavepackets; as a result, $Z_0=\pm 1$ is encoded coherently to two bosonic squeezed states $\ket{\varphi_\pm}$ at position $\hat{x}\sim \pm \sqrt{N}$, corresponding to $X\sim \pm N$. This squeezing-displacement-antisqueezing approach is a standard method to amplify signals \cite{SATIN16,SATIN16_Dur,SATIN_boson19,SATIN_robust17,SATIN_exp22}. We then use the ancillae to control the target data qubits $S$ using Hamiltonian $H^{\rm b}_{\rm DCZ}=\sum_{i\in S}Z_i \otimes V(\hat{x})$, where ``potential'' $V(\hat{x})=\sqrt{2N}\hat{x}+\cdots$ is an odd polynomial of $\hat{x}$ that almost stabilizes the two squeezed states $\ket{\varphi_\pm}$ with an energy difference $\Delta V\sim N$. This imprints a $\frac{\pi}{2}$-phase difference with respect to $Z_iZ_0=\pm 1$ in time $T_{\rm DCZ}\sim 1/N$. 
Finally, we reverse the signal-amplification step and apply on-site rotations to generate the desired CZ gates. Although the protocol contains an error because $V(\hat{x})$ does not stabilize $\ket{\varphi_\pm}$ exactly, it can be made polynomially small by choosing a sufficiently high but constant polynomial degree of $V(\hat{x})$, making it sufficiently flat at the two positions.

Of course, the ancillae are not a true boson mode, but the boson protocol above can be simulated by expanding the bosonic Hamiltonians $H^{\rm b}_{\rm S},H^{\rm b}_{\rm CD},H^{\rm b}_{\rm DCZ}$ using \eqref{eq:HP} and truncating at order $\approx K$, which becomes an all-to-all Hamiltonian \eqref{eq:Hint=}. For example, $H^{\rm b}_{\rm S}=-\frac{1}{4}(XY+YX)+\cdots$ is just the two-axis-twisting spin squeezing Hamiltonian \cite{squeeze93} at leading order of $\frac{N-Z}{2N}$. By controlling the states to never evolve outside the region $\frac{N-Z}{2N}\le N^{-\delta}$ with a small constant $\delta$ determined by $\delta_{\rm T}$, we can bound the simulation error as roughly $N^{-\delta K}=1/\poly{N}$ for a sufficiently large but constant $K$. Furthermore, the qubit protocol takes similar time $T\sim 1/N$ as the bosonic one, which leads to Theorem \ref{thm:CZ}.

\section{$\sqrt{N}$ speedup for any quantum circuit}
Corollary \ref{cor:GHZ_W} shows that certain quantum circuits can be simulated by Hamiltonians $\sim N$-times faster with constant space overhead. However, it is unclear whether this holds for general circuits. Here we show our second main result that any target quantum circuit can be simulated with at least a weaker $\sim \sqrt{N}$ speedup, using $2$-local Hamiltonians and a constant space overhead (see Corollary 6.2 in SM \cite{SM}):

\begin{thm}\label{thm:circ}
    For any constants $0<\delta_{\rm T}<1/2$ and $\kappa>0$, if $N$ is sufficiently large, then any quantum circuit $U_{\rm cir}$ of depth $D$ on the data qubits can be simulated by a Hamiltonian protocol $H(t)$ with $K=2$ \revise{and normalization \eqref{eq:Hint=}} 
    in time \begin{equation}\label{eq:T=DsqrtN}
        T\le \widetilde{c} N^{-\frac{1}{2}+\delta_{\rm T}}D,
    \end{equation}
    where the on-site fields $h_i^a(t)$ are allowed to be random. The simulation error \begin{equation}\label{eq:Epsi<_random}
        \mathbb{E}_{H} \norm{\lr{\mathcal{T}\ee^{-\ii\int^T_0 \dd tH(t)}-U_{\rm cir}}\kpsi\otimes \kzero}^2 \le c D^2 N^{-2\kappa},
    \end{equation}
    is small for any given input state $\kpsi$, averaged over the randomness of the Hamiltonian $\mathbb{E}_{H}$. Here $c,\widetilde{c}$ are constants determined by $\delta_{\rm T},\kappa$. \revise{Moreover, the Hamiltonian norm obeys $\norm{H(t)}\le N^{3/2}$.}
\end{thm}


For any polynomial-depth circuit we can choose a sufficiently large constant $\kappa$ so that the error \eqref{eq:Epsi<_random} is $1/\poly{N}$, which is sufficiently small for meaningful purposes. 
\comment{
Theorem \ref{thm:circ} comes from an interesting version without
randomness, which provides a deterministic simulation in time \eqref{eq:T=DsqrtN} that succeeds for almost but not all inputs $\kpsi$. More precisely, the simulation is accurate for a random $\kpsi$ drawn from a state $1$-design $\mathbb{E}_\psi\kpsi\bra{\psi}\propto I$. This yields an ``optimistic'' version of the target unitary, which could be sufficient for practical purposes like implementing Shor's factoring algorithm \cite{optimistic_circuit25}. More precisely,}
\revise{As an example, Theorem \ref{thm:circ} implies} that Shor's algorithm can be implemented in time $T\sim \sqrt{N}$ using \revise{programmable} all-to-all Hamiltonians with a constant space overhead, \revise{by simulating a $D\sim N$} circuit \cite{optimistic_circuit25}. \revise{With the trapped ion normalization $H_{\rm phys}=N^{-1}H$, this becomes $T_{\rm phys}\sim N^{3/2}$. Although $T_{\rm phys}\gg D$, realizing the circuit requires extra routing overhead; our protocol could provide practical advantage comparing to e.g. 1d local circuits where the smallest known depth for Shor's algorithm is $D=\Theta(N^2)$ \cite{shor_1d} to the best of our knowledge. }

\revise{
\subsection{The fast scrambling conjecture for dense Hamiltonians}

Holographic models of quantum gravity provide a precise setting for the fast scrambling conjecture motivated by black hole physics. In this context, it is natural to use an \emph{extensive} normalization for the Hamiltonian,
\begin{equation}\label{eq:H_extensive}
    \norm{H^{\rm E}(t)}=\order(N).
\end{equation}
With this normalization, the depth of an all-to-all circuit can be compared directly with the evolution time of a Hamiltonian. Recent work \cite{randomU_25} gives an operational proof of the fast scrambling conjecture for all-to-all circuits: circuits of depth $\order(\log N)$ can generate unitary designs \cite{tdesign03} and pseudorandom unitaries (PRUs) \cite{PRU18} that are indistinguishable from Haar-random unitaries in the appropriate finite-query or computational senses. This realizes the physical intuition that information can become essentially fully scrambled on the fastest possible timescale $\order(\log N)$. Moreover, the randomness of the generated ensembles $\{U\}$ is strong: an adversary cannot distinguish them from the Haar ensemble even with query access to $U$ together with its complex conjugate $U^*$, Hermitian conjugate $U^\dagger$, and transpose $U^{\rm T}$.

More realistic holographic models, such as SYK, are denser than all-to-all circuits in the sense that many more pairs interact simultaneously, but each individual coupling is weaker. For example, consider the Sachdev-Ye (SY) model \cite{SYK93} $H^{\rm SY}=\frac{1}{\sqrt{N}}\sum_{ij}\sum_a J_{ij}X_i^aX_j^a$
where $J_{ij}=\order(1)$ are independent random variables. It satisfies the extensive normalization \eqref{eq:H_extensive} with high probability, while every qubit is coupled to almost all other qubits and each individual coupling obeys $\norm{H^{\rm SY}_{ij}}=\order(N^{-1/2})$. We say that a $2$-local Hamiltonian $H^{\rm E}(t)=\sum_{ij}H^{\rm E}_{ij}(t)$ is $\fkd$-dense if
\begin{equation}\label{eq:dense}
    \max_{t,ij}\norm{H^{\rm E}_{ij}(t)} \le N^{-\fkd}.
\end{equation}
Thus the SY model is $1/2$-dense, whereas all-to-all circuits correspond to the endpoint $\fkd=0$.
It is then natural to ask whether dense Hamiltonian evolutions with $\fkd>0$ can still generate unitary designs or PRUs in nearly logarithmic time under the extensive normalization \eqref{eq:H_extensive}. As a corollary of Theorem \ref{thm:circ}, applying our simulation protocol to the circuit ensembles of \cite{randomU_25} gives an affirmative answer up to a subpolynomial-time overhead; see Corollaries 7.2 and 7.4 of the SM \cite{SM}:

\begin{cor}[Informal]\label{cor:randomU}
    For any constant $\delta>0$ and any $0<\fkd\le 1/2$, for all sufficiently large $N$ there exists an ensemble of Hamiltonian evolutions $H^{\rm E}(t)$ on $\order(N)$ qubits satisfying \eqref{eq:H_extensive} and \eqref{eq:dense} such that a strong unitary $k$-design ensemble on $N$ qubits, with superpolynomially small measurable error, is generated in time
    \begin{equation}\label{eq:TE<N}
        T^{\rm E}\le N^\delta
    \end{equation}
    for all $k\le N^{\delta/2}$. Under standard cryptographic assumptions, an ensemble of Hamiltonian evolutions generating a strong PRU ensemble on $N$ qubits also exists within the same time \eqref{eq:TE<N}.
\end{cor}

Although Corollary \ref{cor:randomU} does not yet give a logarithmic-time guarantee, the bound \eqref{eq:TE<N} is faster than any fixed power of $N$. We therefore view the remaining gap to $\polylog(N)$ time as largely technical. Compared with the $\polylog(N)$-depth circuits of \cite{randomU_25}, which do not use ancillae, Corollary \ref{cor:randomU} is still unsatisfactory because it uses $\order(N)$ ancilla qubits. Removing or reducing this ancilla overhead is an interesting open problem.
}

\subsection{Proof sketch}

\begin{figure}
    \centering
    \includegraphics[width=0.8\linewidth]{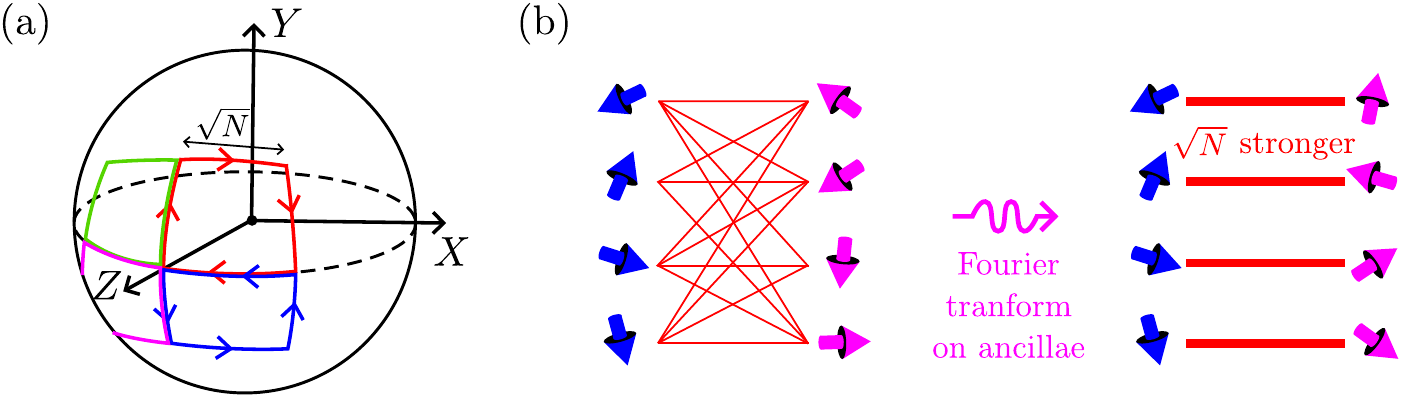}
    \caption{Sketch of the idea for Theorem \ref{thm:circ}. (a) An exact $T=\order(1/\sqrt{N})$ protocol of simulating one gate $\mathrm{CZ}_{0,1}$. The four colored trajectories correspond to the four choices of $Z_0,Z_1=\pm1$. 
    (b) We Fourier transform the ancilla qubits $X_i^+\rightarrow \widetilde{X}_k^+$, which focus the coupling strengths to make them individually stronger. Although the Fourier modes $\widetilde{X}_k^+$ are not simple spins or bosons, they can be approximated so for almost all inputs of our simulation protocol.
    }
    \label{fig:sqrtN}
\end{figure}

\revise{Here we sketch the proof ideas behind Theorem \ref{thm:circ}. First, we show that there exists} a $T=\order( 1/\sqrt{N})$ \emph{exact} protocol for a single $\mathrm{CZ}$ gate (Theorem 6.4 in SM \cite{SM}). Although it is slower than the approximate one in Theorem \ref{thm:CZ}, we illustrate its idea here for later generalization to Theorem \ref{thm:circ}. As Theorem \ref{thm:CZ}, the $N$ ancilla qubits work in their Dicke manifold with semiclassical angular momentums $X_{\rm cl}=X/N,Y_{\rm cl},Z_{\rm cl}$. We mimick the Mølmer-Sørensen scheme \cite{MS99,MS99_} that mediates qubit-qubit interactions by coupling them to a boson mode. The protocol contains four stages where $H(t)=\pm Z_0\otimes X$ for the first and third stage while $H(t)=\pm Z_1\otimes Y$ for the second and fourth, where $0,1$ are the two data qubits of the CZ gate. As shown in Fig.~\ref{fig:sqrtN}(a), $H(t)$ rotates the initial $\kzero$ to spin coherent states at different solid angles, controlled by the data qubits. By tuning the time duration $T_\mu=\order(1/\sqrt{N})$ for each stage $\mu$, the trajectory returns to $\kzero$ exactly and decouples the ancillae. Furthermore, the trajectory encloses a phase space area with $A/\hbar=\Theta(1)$ where $A\sim \Delta X_{\rm cl} \Delta Y_{\rm cl}=\Theta(1/N)$ is the action and $\hbar=1/N$. As a result, for a particular choice of $T_\mu=\order(1/\sqrt{N})$, the data qubits gain an exact $\frac{\pi}{2}$ geometric phase depending on whether $Z_0Z_1=\pm 1$. This yields an exact $\mathrm{CZ}_{0,1}$ with on-site rotations.

We then turn to Theorem \ref{thm:circ}, which boils down to a protocol that simulates one layer of CZ gates of the target circuit in $T\sim 1/\sqrt{N}$. In other words we need to parallelize the exact single-CZ protocol to simulate $\order(N)$ gates. The idea is to ``focus'' the $\mathrm{\Theta}(N^2)$ couplings of strength $\order(1)$ in the $2$-local Hamiltonian to $\mathrm{\Theta}(N)$ couplings of stronger strength $\mathrm{\Theta}(\sqrt{N})$, as shown in Fig.~\ref{fig:sqrtN}(b). If the $N$ ancilae were bosons $\{b_i\}$, this is achievable by Fourier transforming the boson modes $b_i \rightarrow \widetilde{b}_k$ so that each data qubit $k$ (together with its companion qubit in the same CZ gate) is coupled to its own Fourier boson: $Z_k\otimes \sum_i \ee^{\ii \frac{2\pi k}{N}i} b_i = \sqrt{N} Z_k\otimes \widetilde{b}_k$. Note that this $\sqrt{N}$ enhancement also underlies a recent fermionic routing protocol \cite{our_routing25}. We then apply the Mølmer-Sørensen scheme for each $\{2$ data qubits, $1$ Fourier boson$\}$ pair that is $\sqrt{N}$-times faster than coupling the qubits to original bosons individually. We then obtain a qubit protocol by replacing each $b_i$ by their counterpart $X^+_i=X_i+\ii Y_i$ in \eqref{eq:HP}\revise{:
\begin{equation}\label{eq:A=sumki}
    H(t)=\ee^{\ii \phi(t)}\sum_{k,i} \ee^{\ii \frac{2\pi k}{N}i}Z_k\otimes (X_i+\ii Y_i) + \mathrm{H.c.} =: \sum_k Z_k\otimes \widetilde{X}^+_k,
\end{equation}
where $k$ and $i$ label data and ancilla qubits, respectively. The time dependence of $H(t)$ comes from the global phase $\phi(t)$ and from permutations of the data-qubit labels $k$. One can verify that $\norm{H(t)}=\order(N^{3/2})$, which is crucial for Corollary \ref{cor:randomU}. } Although the Fourier transformations of qubits $\widetilde{X}^+_k$ are neither spin nor boson operators anymore, it turns out that during the protocol each ancilla rarely visits its ``excited'' state $\ket{1}$ 
for \emph{almost} all inputs. This can be seen from the single-CZ protocol in Fig.~\ref{fig:sqrtN}(a), where the dynamics remains close to the north pole. Although here each ancilla qubit is coupled to $\Theta(N)$ data qubits $j$ instead of one, the rotation angle $\theta = T \sum_j Z_j=\order(N^{-\delta_{\rm T}})\ll 1$ from $T=N^{-\frac{1}{2}+\delta_{\rm T}}$ for a typical choice of $\{Z_j\}$. Since a qubit feels the difference between $X^+_i$ and $b_i$ only when it is excited, which happens with small probability $\order(N^{-\delta_{\rm T}})$ during the protocol, the Fourier enhancement trick for bosons above also works for qubits, leading to a protocol in time \eqref{eq:T=DsqrtN} with error $\propto N^{-\delta_{\rm T}}$ for almost all input $\kpsi$. 

By generalizing the Mølmer-Sørensen scheme to higher orders using Suzuki's method \cite{suzuki91}, we are able to bootstrap the error to any inverse-polynomial $1/\poly{N}$. Finally, we apply a standard method of worst-case to average-case reductions \cite{optimistic_circuit25} to get the randomized protocol Theorem \ref{thm:circ} that works for any input. The idea is to apply random on-site rotations to deliberately randomize the input state before the simulation, and undo the random rotations afterwards.

\section{Discussion and Outlook}\label{sec:conclude}
We have proven that all-to-all Hamiltonians can process information much faster than all-to-all quantum circuits, assuming comparable qubit-qubit interaction strength for the two models. Although quantum computation is well established for the circuit model, all-to-all Hamiltonians could be a more natural model to consider for experimental \cite{cavity_metro_scram23,cavity_graphstate24,ion_longrange12,ion_rev21,all2all_supercond} and physical \cite{Dickemodel54,LMG65,SK75,SK_Parisi79,SYK93,SYK_talk15,SYK_remark16} settings. \revise{In particular, our fast protocols in such dense Hamiltonians provide new mathematical evidence on the fast scrambling conjecture.} Our work pioneers the study of complexity in the Hamiltonian model, where we show that one typically needs to apply ideas drastically different from circuits, like squeezing from non-commuting Hamiltonians. 

\revise{Regarding experimental realizations, our protocols are potentially useful for reducing routing overheads. Although a fully programmable Hamiltonian with complicated time dependence\footnote{See \cite{tindepen_free25,tindepen_clock25,all2all_tindepen22,all2all_tindepen25} for existing time-independent constructions.} and inhomogeneous couplings $H_{ij}$ may be challenging to realize}, we expect simple versions of our ideas, like \cite{myGHZ}, could be tangible and enable much faster protocols to process quantum information. \revise{Intriguingly, Fourier-transform-like coupling coefficients of the form appearing in \eqref{eq:A=sumki} arise naturally in confocal multimode cavities \cite{cavity_Jij_vector}. Since \eqref{eq:A=sumki} is one ingredient in our Hamiltonian realization of fast scrambling, simulating Hamiltonians with this structure may offer a more accessible route to experimentally probing aspects of quantum-gravity-inspired scrambling than realizing a full SYK model with independently random couplings. One possible route is the following. Start from a confocal-cavity building block $H_0=\sum_{ki}J_{ki}X_iX_k$ (see Eq.~(5) of \cite{lev_theory24}), where $J_{ki}=\cos(\alpha \bm{r}_k\cdot \bm{r}_i)$ for a constant $\alpha$ and position vectors $\bm{r}_i$. Using two pump configurations, one can engineer a combination $H_0+H'_0$ in which the qubits are divided into two groups, $\{k\}$ and $\{i\}$, and the intra-group couplings cancel, leaving primarily inter-group couplings. Dressing the qubits by on-site rotations can then convert these interactions into $ZX$ and $ZY$-type couplings. Finally, place the two groups in one-dimensional arrays with positions $\bm{r}_k=(\beta k,y_1,0)$ and $\bm{r}_i=(\beta i,y_2,0)$. The kernel of $J_{ki}$ then contains a phase $\alpha\beta^2ki$ plus a tunable offset from $y_1y_2$, so that by choosing two values of $y_2$ one can obtain the cosine and sine components needed in \eqref{eq:A=sumki}. In principle, \eqref{eq:A=sumki} could then be approximated by Floquet engineering if the motion between the two values of $y_2$, together with the corresponding switch of the $X,Y$ basis, is fast compared with the relevant interaction timescale. Since experimental capabilities such as cavity cooperativity continue to improve, and better candidate Hamiltonians motivated by \eqref{eq:A=sumki} may be further identified, we leave a detailed feasibility analysis of these proposals to future work.}

Our work also opens up interesting theoretical questions. \revise{In local systems, Lieb-Robinson bounds underlie the universality of gapped ground-state phases \cite{quasiadiabatic05}: different states in the same phase are connected by quasi-local unitaries generated by a quasi-local Hamiltonian $\widetilde{H}(t)$, with evolution time $T=\order(1)$ set by the $\Omega(1)$ gap. Our fast protocols raise a natural question about how this picture should be formulated in strongly long-range systems, including fractional quantum Hall states in a two-dimensional electron gas with Coulomb interactions $J_{ij}\sim r_{ij}^{-1}\gg r_{ij}^{-2}$. In standard quasi-adiabatic evolution \cite{quasiadiabatic05}, the generator $\widetilde{H}$ inherits long-range tails from the physical Hamiltonian $H$ and is generally noncommuting even when the long-range part of $H$ is commuting. According to Corollary \ref{cor:LRB}, Hamiltonians of this type can propagate information globally in a time $T\ll 1$ that vanishes with system size. This suggests two possibilities: either our fast protocols are highly fine-tuned, so that the quasi-adiabatic generators arising in physical gapped phases have additional structure that still preserves an appropriate notion of locality, or the standard quasi-adiabatic framework must be refined to describe universality in strongly long-range gapped phases. Either outcome would be novel compared with geometrically local physics; determining which alternative holds remains an open question.}

\revise{It would also be interesting to determine whether some of our fast protocols can be further improved, such as the $\sqrt{N}$-speedup for arbitrary circuits or the $1/N$-time protocol for preparing the W state.}
Perhaps more interestingly, there could exist ways of processing information that are intrinsic to Hamiltonians and do not come from simulating any circuit. It is also unclear how to perform error correction if noise interrupts Hamiltonian evolution. We believe these questions could inspire fruitful investigations connecting many-body physics, computer science and quantum technology.

\emph{Acknowledgement.}---
We thank Andrew Childs, Andrew Lucas, Zhengyan Darius Shi and Haoqing Zhang for helpful discussions. We thank Peter McMahon and Hongzheng Zhao for drawing our attention to relevant references.
This material is based upon work partially supported by the National Science Foundation under award No. 2016245 and by the Stanford Q-FARM Bloch Postdoctoral Fellowship in Quantum Science and Engineering.

\bibliography{biblio}



\section*{Supplementary material}

# Supplementary Material: Fast quantum computation with all-to-all Hamiltonians

Chao Yin*

*Department of Physics, Stanford University, Stanford, California 94305, USA*

(Dated: September 29, 2025)

**CONTENTS**

―――――

* chaoyin@stanford.edu

# 1. PRELIMINARIES

## 1.1. All-to-all Hamiltonians

We consider a total number of $|\Lambda| = N_{\text{tot}}$ qubits labeled by $i \in \Lambda \subset \mathbb{Z}$. We use two notations for Pauli matrices on $i$: $X_i \equiv X_i^1, Y_i \equiv X_i^2, Z_i \equiv X_i^3$. Suppose one can engineer any time-dependent Hamiltonian of the form

$$H(t) = H_{\text{rot}}(t) + H_{\text{int}}(t), \tag{1.1}$$

where $H_{\text{rot}}(t) = \sum_i h_i^a(t) X_i^a$ is single-qubit rotations with arbitrary field strength $h_i^a$ (we use Einstein summation rules for index $a = 1, 2, 3$), and the interaction part

$$H_{\text{int}}(t) = \sum_{k=2}^{K} N_{\text{tot}}^{2-k} \sum_{i_1 < i_2 < \cdots < i_k} J_{i_1 \cdots i_k}^{a_1 \cdots a_k}(t) X_{i_1}^{a_1} \cdots X_{i_k}^{a_k}, \quad \text{where} \quad \sum_{a_1 \cdots a_k} \left| J_{i_1 \cdots i_k}^{a_1 \cdots a_k}(t) \right| \leq 1. \tag{1.2}$$

The Hamiltonian $H_{\text{int}}(t)$ is $K$-local; the special case $K = 2$ reduces to the usual 2-local Hamiltonians like $H = \sum_{ij} J_{ij} Z_i Z_j$ with $J_{ij} \sim 1$, i.e. each pair of qubits interact with O(1) strength. The $N_{\text{tot}}^{2-k}$ factor in (1.2) makes $k$-local terms with different $k$ have comparable total interaction strength, so that

$$\|H_{\text{int}}(t)\| \leq \sum_{k=2}^{K} N_{\text{tot}}^{2-k} \binom{N_{\text{tot}}}{k} \leq N_{\text{tot}}^2 \sum_{k=2}^{K} \frac{1}{k!} \leq N_{\text{tot}}^2, \tag{1.3}$$

where we have used $\left\| X_{i_1}^{a_1} \cdots \right\| = 1$ and triangle inequality. Such a $N$ scaling difference among different $k$s is a natural assumption; see discussions in Section 7.

Evolving the system in time $T$ generates unitary

$$U = \mathcal{T} e^{-i \int_0^T dt\, H(t)}, \tag{1.4}$$

where $\mathcal{T}$ is time-ordering. We assume $T > 0$ throughout the manuscript.

## 1.2. Quantum circuits

We say $U_{\text{cir}}$ is a depth-$D$ quantum circuit, if $U_{\text{cir}} = U_{\text{rot},D} U_{\text{cir},D} \cdots U_{\text{rot},2} U_{\text{cir},2} U_{\text{rot},1} U_{\text{cir},1}$ where each layer is either single-qubit rotations $U_{\text{rot},m}$, or a bunch of controlled-Z (CZ) gates

$$U_{\text{cir},m} = \prod_{j \in S} \text{CZ}_{j,\tau(j)}, \tag{1.5}$$

where different pairs of qubits $(j, \tau(j))$ ($\tau(j)$ being the "target" qubit of the "control" qubit $j$) do not overlap, and $S \subset \Lambda$ is the set of control qubits. A single CZ gate is $\text{CZ}_{i,j} := I - 2 \left| 11 \right\rangle_{i,j} \left\langle 11 \right|$. A quantum circuit with more general two-qubit gates can be written in the form above with a mild overhead $D \to 3D$, because any two-qubit gate can be decomposed to on-site rotations and at most 3 controlled-NOT gates that are equivalent to CZ [1].

Any depth-$D$ quantum circuit composed of single- and two-qubit gates can be simulated by $H(t)$ in time $T = \mathrm{O}(D)$, by turning on the 2-local coupling between two qubits when they are involved in a two-qubit gate. On the other hand, known algorithms for simulating $H(t)$ by quantum circuits require a polynomial poly $(N)$ overhead in spacetime [2, 3]. The goal of this manuscript is to show that $H(t)$ indeed can achieve much faster quantum computation comparing to the naive way of simulating circuits: First, certain interesting families of circuits and states can be generated faithfully in time $T \sim D/N_{\text{tot}}$. Second, any circuit can be simulated by a weaker speed up $T \sim D/\sqrt{N_{\text{tot}}}$.

## 1.3. Other notations and structure of manuscript

We introduce several notations. We use $N$ of the total $N_{\text{tot}}$ qubits $i = 1, 2, \cdots, N$ as ancillae, which always start the evolution from the all-zero state $|\mathbf{0}\rangle$, and approximately returns to $|\mathbf{0}\rangle$ in the end of the evolution. Throughout the manuscript, we assume $N$ is sufficiently large: $N > c$ for a constant $c$ determined by the other constants like $K$. For simplicity, however, we do not calculate $c$ explicitly. We will say $c'$ is a constant if it does not depend on $N$.

We use the big-O notation $y = \mathrm{O}(x)$ to mean $|y| \leq cx$ for some constant $c$. $y = \Omega(x), y = \Theta(x)$ mean $|y| \geq cx$ and $c'x \leq |y| \leq cx$ similarly. We use $y \sim x$ to mean rough estimates that $y$ and $x$ are of same order; $y/x$ could be $N^{\pm\delta}$ with a small constant $\delta$.

The manuscript is structured as follows. In Section 2 we develop some technical tools on mapping the ancilla qubits to a boson mode, which underlie our results[1]. In Section 3, we present our first main result on simulating any two-qubit gate in $\sim 1/N$ time, which generalizes the Greenberger-Horne-Zeilinger (GHZ) encoding protocol in [4] so that it can be rigorously analyzed. The proof is rather technical, with many details arranged in Section 4. We discuss implications of this protocol in Section 5, including for example interesting families of circuits aided by unbounded fan-out gates simulable in $\sim 1/N$ time, and Lieb-Robinson bounds in power-law interacting systems that match our protocol. In Section 6, we present our second main result on the $\sqrt{N}$ speed up for simulating any circuit. An implication is that Shor's algorithm [5] can be simulated by Hamiltonians in time $T \sim \sqrt{N}$ with a small space overhead. Finally in Section 7, we discuss our assumptions and possible extensions on the Hamiltonian form we use.

## 2. MAPPING AN ENSEMBLE OF QUBITS TO A BOSON MODE

### 2.1. Holstein-Primakoff transformation

In this section we focus on the $N$ ancilla qubits. The global angular momentums are $X = \sum_{i=1}^{N} X_i$ and $Y, Z$ defined similarly, which form a SU(2) algebra. The Hilbert space factorizes into the irreducible representations (IRREP) of the SU(2), where the largest IRREP of dimension $N + 1$ is known as the Dicke manifold (DM) of the $N$ qubits.

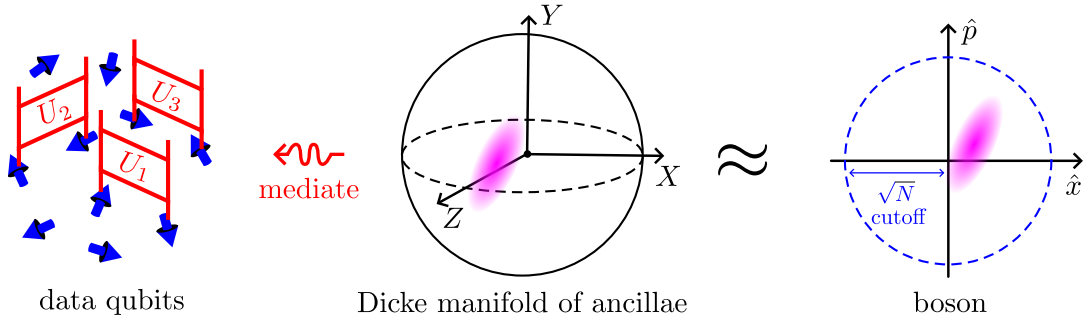

FIG. 1. For our first main result Theorem 3.1, we mediate interactions (gates) on data qubits by $N$ ancila qubits, which work in their DM that resembles a boson mode close to its north pole $Z = N$. A boson state is well inside the DM Hilbert space as long as its boson number $B^\dagger B \ll N$, or equivalently, the distance to origin in phase space is $\ll \sqrt{N}$. Our protocol works in this regime and simulates boson Hamiltonians, e.g. squeezing $\mathrm{i}(B^2 - B^{\dagger 2})$, by all-to-all qubit Hamiltonians.

In most of our protocols (except Theorem 6.1), the dynamics of the ancillae is restricted to the DM, which can be viewed as a large semiclassical spin with total angular momentum $X^2 + Y^2 + Z^2 = N(N + 2)^2$ and a spherical phase space, as shown in Fig. 1. Close to the north pole of the sphere, this large spin resembles a boson mode by the Holstein-Primakoff (HP) transformation (see e.g. textbook [6])

$$Z = N - 2b^\dagger b, \tag{2.1a}$$

$$X^+ = 2\sqrt{N - b^\dagger b}\, b, \tag{2.1b}$$

$$X^- = 2b^\dagger \sqrt{N - b^\dagger b}, \tag{2.1c}$$

where $X^\pm = X \pm \mathrm{i}Y$ satisfies $[X^+, X^-] = 4Z$. The inverse transformation is

$$b = \frac{1}{2\sqrt{N}} \left(1 - \frac{N - Z}{2N}\right)^{-1/2} X^+. \tag{2.2}$$

---

Note that here

$$b = PBP = BP, \tag{2.3}$$

is not a true boson annihilation operator, but one ($B$) truncated in the boson number $\leq N$ subspace, i.e. the physical Hilbert space of the DM whose projection is denoted by $P$. For example we have

$$[B, B^\dagger] = 1, \tag{2.4}$$

but $[b, b^\dagger] \neq 1$ with violation only at the truncation.

We will frequently use identity (2.4) and

$$f(\hat{n})B = Bf(\hat{n} - 1), \quad B^\dagger f(\hat{n}) = f(\hat{n} - 1)B^\dagger \tag{2.5}$$

to simplify operators, where $f(\hat{n})$ is any operator that is a function of boson number

$$\hat{n} := B^\dagger B, \tag{2.6}$$

(i.e. diagonal operator in boson number basis). Another way to think about a boson is based on the phase space with position and momentum

$$\hat{x} = \frac{B + B^\dagger}{\sqrt{2}}, \quad \hat{p} = \frac{B - B^\dagger}{\sqrt{2}i}. \tag{2.7}$$

A bosonic state can be represented by a distribution (e.g. Wigner distribution) in phase space, which we refer to as a wavepacket.

We will consider dynamics in the enlarged Hilbert space, which will be approximated by the real physical protocol generated by $H(t)$ that preserves subspace $P$.

### 2.2. Bosonic squeezed states

Let $|0\rangle$ be the $\hat{n} = 0$ state, i.e. the all-zero state of the qubits. We review properties of the displaced squeezed state (see e.g. [7])

$$|\zeta, \alpha\rangle := e^{\alpha B^\dagger - \alpha^* B} e^{\frac{\zeta}{2}(B^2 - B^{\dagger 2})} |0\rangle, \tag{2.8}$$

with complex parameter $\alpha$ and real $\zeta$, which is also known as a coherent state. One can easily define more general states with complex $\zeta$ as well, but real number $\zeta$ is sufficient for our purpose and simplifies the expressions. When $\zeta = 0$ the state can be expanded in the boson number basis as

$$|0, \alpha\rangle = e^{-\frac{|\alpha|^2}{2}} \sum_n \frac{\alpha^n}{\sqrt{n!}} |n\rangle, \tag{2.9}$$

where $\hat{n} |n\rangle = n |n\rangle$. Similarly for $\alpha = 0$ and $\zeta$ real,

$$|\zeta, 0\rangle = \frac{1}{\sqrt{\cosh \zeta}} \sum_{n=0}^\infty (-\tanh \zeta)^n \frac{\sqrt{(2n)!}}{2^n n!} |2n\rangle. \tag{2.10}$$

Define $\langle \mathcal{O} \rangle_{\zeta,\alpha} := \langle \zeta, \alpha | \mathcal{O} | \zeta, \alpha \rangle$. The displaced squeezed state is solvable because its generating Hamiltonian is quadratic in $B, B^\dagger$. More precisely, it satisfies Wick's theorem:

**Proposition 2.1** (Wick's theorem). *Let $\mathbb{B}_1, \mathbb{B}_2, \cdots$ be either $B - \alpha$ or $B^\dagger - \alpha^*$.*

$$\langle \mathbb{B}_1 \mathbb{B}_2 \mathbb{B}_3 \cdots \rangle_{\zeta,\alpha} = \langle \mathbb{B}_1 \mathbb{B}_2 \rangle_{\zeta,\alpha} \langle \mathbb{B}_3 \mathbb{B}_4 \rangle_{\zeta,\alpha} \cdots + \langle \mathbb{B}_1 \mathbb{B}_3 \rangle_{\zeta,\alpha} \langle \mathbb{B}_2 \mathbb{B}_4 \rangle_{\zeta,\alpha} \cdots + \cdots, \tag{2.11}$$

*expands into all possible pairwise contractions. Each pairwise contraction is*

$$\langle B - \alpha \rangle_{\zeta,\alpha} = 0, \tag{2.12a}$$

$$\left\langle (B - \alpha)^2 \right\rangle_{\zeta,\alpha} = -\cosh(\zeta)\sinh(\zeta), \tag{2.12b}$$

$$\left\langle (B - \alpha)(B^\dagger - \alpha^*) \right\rangle_{\zeta,\alpha} - 1 = \left\langle (B^\dagger - \alpha^*)(B - \alpha) \right\rangle_{\zeta,\alpha} = \sinh^2(\zeta), \tag{2.12c}$$

*and their complex conjugates.*

## 2.3. Truncation errors

Recall that projector $P$ truncates the boson number to $\leq N$. More generally, we define projector $P_{<n}$ ($P_{>n}$) as truncating into the $\hat{n} < n$ ($\hat{n} > n$) subspace. We will also use $P_{\mathcal{O}>n}$ to represent the projector onto the eigensubspaces of operator $\mathcal{O}$ with eigenvalue $\lambda > n$. Similarly, $P_{|\mathcal{O}|>n}$ is for the subspace $\mathcal{O} > n$ or $\mathcal{O} < -n$.

We show that the boson number concentrates in coherent states:

**Lemma 2.2.** *For real* $\alpha$,

$$\left\| P_{|\hat{n} - \alpha^2| > \Delta} \, |\zeta, \alpha\rangle \right\| \leq 2 \exp\left[1 - \frac{0.6}{e^{|\zeta|}|\alpha|} \Delta\right], \tag{2.13}$$

*for all*

$$|\alpha| > 5 e^{|\zeta|}. \tag{2.14}$$

The physical intuition is that in the boson phase space, state $|n\rangle$ is supported at a circle centered at the origin of radius $\Theta(\sqrt{n})$, while $|\zeta, \alpha\rangle$ is supported near complex coordinate $\alpha$ if not too squeezed (2.25a). This coherent state thus overlaps with only the $|n\rangle$s whose circles go nearby $\alpha$. See Fig. 1 for an illustration.

*Proof.* We use the generating function $G(\lambda) := \left\langle \lambda^{\hat{n}} \right\rangle_{\zeta,\alpha}$ for boson number derived in Eq.(28) of [8] (see also [9])

$$G(\lambda) = |\langle 0|\zeta, \alpha\rangle|^2 \left[\left(1 - \frac{\lambda}{\lambda_1}\right)\left(1 - \frac{\lambda}{\lambda_2}\right)\right]^{-1/2} \exp\left[\frac{\lambda x_1}{\lambda - \lambda_1} + \frac{\lambda x_2}{\lambda - \lambda_2}\right], \tag{2.15}$$

where

$$|\langle 0|\zeta, \alpha\rangle|^2 = \frac{1}{\cosh\zeta} e^{-\alpha^2(1+\tanh\zeta)}, \tag{2.16a}$$

$$\lambda_1 = -\lambda_2 = \frac{2\sinh(2\zeta)}{2 + 2\cosh(2\zeta)} = \tanh\zeta, \tag{2.16b}$$

$$y_1 = \alpha(1 + \coth\zeta), \tag{2.16c}$$

$$x_1 = y_1^2/\tanh\zeta = \alpha^2(1 + \tanh\zeta)(1 + \coth\zeta), \tag{2.16d}$$

$$x_2 = 0, \tag{2.16e}$$

for our case of real $\alpha, \zeta$.

Plugging these into (2.15) yields

$$\begin{aligned} G(\lambda) &= \frac{1}{\cosh\zeta} \frac{1}{\sqrt{1 - \lambda^2 \tanh^2\zeta}} \exp\left[\alpha^2(1+\tanh\zeta)\left(\frac{\lambda(1+\tanh\zeta)}{1 + \lambda\tanh\zeta} - 1\right)\right] \\ &= \frac{1}{\cosh\zeta} \frac{1}{\sqrt{1 - \lambda^2 \tanh^2\zeta}} \exp\left[\alpha^2(\lambda-1)\left(1 - \frac{\lambda - 1}{\lambda + \coth\zeta}\right)\right] \tag{2.17} \\ &\leq \frac{1}{\cosh\zeta} \frac{1}{\sqrt{1 - \lambda^2 \tanh^2\zeta}} \exp\left[\alpha^2\left(\lambda - 1 + e^{2|\zeta|}(\lambda-1)^2\right)\right], \tag{2.18} \end{aligned}$$

for

$$\frac{1}{2} \leq \lambda < |\coth\zeta| - e^{-2|\zeta|}. \tag{2.19}$$

For case $n > \alpha^2$, we choose

$$\lambda - 1 = e^{-|\zeta|}/|\alpha| < 1/5, \tag{2.20}$$

obeying (2.19) due to (2.14). By Markov's inequality, (2.18) leads to

$$\begin{aligned} \|P_{>n}|\zeta, \alpha\rangle\|^2 &\leq G(\lambda)\lambda^{-n} \leq G(\lambda)\exp\left[-\left(\lambda - 1 - (\lambda-1)^2\right)n\right] \\ &\leq \frac{1}{\cosh\zeta} \frac{1}{\sqrt{1 - (1 + e^{-2|\zeta|}/5)^2\tanh^2\zeta}} \exp\left[-(\lambda-1)(n-\alpha^2) + 2 + (\lambda-1)^2\left(n - \alpha^2\right)\right] \\ &\leq 2\exp\left[2 - \frac{1.2}{e^{|\zeta|}|\alpha|}|n - \alpha^2|\right] \tag{2.21} \end{aligned}$$

where we have used $\log \lambda \geq \lambda - 1 - (\lambda - 1)^2$ in range (2.19) to bound $\lambda^{-n}$, and (2.20).

For the small $n$ case, we choose

$$\lambda - 1 = -\mathrm{e}^{-|\zeta|}/|\alpha|, \tag{2.22}$$

instead. One can verify that the analogous (2.21) still holds and $\|P_{<n}|\zeta,\alpha\rangle\|^2$ obeys the same bound in (2.21). Doubling (2.21) and taking square root yields (2.13). □

For more general cases beyond (2.14), we show the following analogous result on truncation error at sufficiently large boson number. In particular, we allow a fairly general operator $\mathcal{O}$ that probes the truncation error:

**Lemma 2.3.** *Fix a constant $\delta \in (0,1)$, an integer $d \geq 0$, and suppose $n \geq c_{\delta,d}$ is sufficiently large comparing to a constant determined by $\delta, d$. Any operator $\mathcal{O}$ that satisfies*

$$\mathcal{O}^\dagger \mathcal{O} = f(\hat{n}), \quad \text{where} \quad f(n') \leq c_d^2 (n')^{2d}, \quad \forall n' > n, \tag{2.23}$$

*with $c_d > 0$ independent of $n'$, has an exponentially small truncation error*

$$\|\mathcal{O}P_{>n}|\zeta,\alpha\rangle\| \leq 2c_d\,\mathrm{e}^{-n^\delta/10}, \tag{2.24}$$

*for all*

$$\mathrm{e}^{-2|\zeta|} \geq 2n^{-(1-\delta)}, \tag{2.25a}$$

$$|\alpha| \leq \sqrt{n/5}. \tag{2.25b}$$

*Proof.* We use the same generating function $G(\lambda)$ (2.17) with a different choice

$$\lambda = \frac{1}{2}\min\left(4, 1 + |\coth\zeta|\right) \geq 1 + \mathrm{e}^{-2|\zeta|} \geq \exp\left(\frac{1}{2}\mathrm{e}^{-2|\zeta|}\right), \tag{2.26}$$

so that

$$G(\lambda) \leq \frac{1}{\cosh\zeta}\frac{1}{\sqrt{1-(1+|\tanh\zeta|)^2/4}}\exp\left[\alpha^2(\lambda-1)(1+1)\right] \leq \frac{4}{\sqrt{3}}\exp\left[2\alpha^2(\lambda-1)\right]. \tag{2.27}$$

We treat our goal by

$$\begin{aligned}
\|\mathcal{O}P_{>n}|\zeta,\alpha\rangle\|^2 &= \sum_{n'>n} f(n')\,|\langle n'|\zeta,\alpha\rangle|^2 \leq \sum_{n'>n} c_d^2(n')^{2d}\,|\langle n'|\zeta,\alpha\rangle|^2 \\
&\leq c_d^2 \sum_{n'>n} \sqrt{\lambda}^{n'}\,|\langle n'|\zeta,\alpha\rangle|^2 \leq c_d^2 G(\lambda)\sqrt{\lambda}^{-n},
\end{aligned} \tag{2.28}$$

where to get the second line we have used

$$\sqrt{\lambda}^{n'} \geq \exp\left(\frac{1}{4}\mathrm{e}^{-2|\zeta|}n'\right) \geq \exp\left(\frac{1}{2}n^{-(1-\delta)}n'\right) \geq (n')^{2d}, \tag{2.29}$$

from (2.26), (2.25a) and the sufficient largeness of $n' > n$.

Combining (2.27) and (2.28) with $\log\lambda \geq \frac{1}{2}(\lambda-1)$ for $\lambda \leq 2$ (2.26), we get

$$\begin{aligned}
\|\mathcal{O}P_{>n}|\zeta,\alpha\rangle\|^2 &\leq c_d^2\frac{4}{\sqrt{3}}\exp\left[-\frac{1}{2}(\lambda-1)(n-4\alpha^2)\right] \\
&\leq c_d^2\frac{4}{\sqrt{3}}\exp\left[-\frac{1}{2}\mathrm{e}^{-2|\zeta|}(n-4\alpha^2)\right] \leq c_d^2\frac{4}{\sqrt{3}}\exp\left[-n^{-(1-\delta)}\frac{n}{5}\right],
\end{aligned} \tag{2.30}$$

using (2.25), which leads to (2.24). □

## 3. FAST TWO-QUBIT GATE IN $1/N$ TIME

**Theorem 3.1.** *Consider a system of $N_{\mathrm{tot}}$ qubits where $N$ of them are ancilla qubits set initially to the all-zero state $|\mathbf{0}\rangle$, and the others are data qubits. For any constants $\delta_{\mathrm{T}} \in (0,1)$, $c_{\mathrm{tot}} \geq 1$ and locality $K \geq 2$, the following holds for sufficiently large $N$ and*

$$N_{\mathrm{tot}} \leq c_{\mathrm{tot}} N. \tag{3.1}$$

*For two data qubits $0, -1$, there exists a protocol $H(t)$ in (1.1) that simulates $\mathrm{CZ}_{0,-1}$ with polynomially small error*

$$\epsilon = N^{2 - \delta_{\mathrm{T}}(\sqrt{K}-1)/2}, \tag{3.2}$$

*in total evolution time*

$$T \leq N^{-1+\delta_{\mathrm{T}}}. \tag{3.3}$$

Here by simulation error $\epsilon$ for a target unitary $U_{\mathrm{tar}}$, we mean

$$\left\| \left( \mathcal{T} \mathrm{e}^{-\mathrm{i}\int_0^T \mathrm{d}t\, H(t)} - U_{\mathrm{tar}} \right) |\psi\rangle \otimes |\mathbf{0}\rangle \right\| \leq \epsilon, \tag{3.4}$$

for any initial state $|\psi\rangle$ on the data qubits. In other words, for any polynomially small error $N^{-\kappa}$ with constant $\kappa$, the CZ gate can be simulated in almost $N^{-1}$ time as long as $K \gg \kappa^2$. We need condition (3.1) because $\geq$ 3-body interactions are essential for our protocol, which will be suppressed by normalization (1.2) if $N_{\mathrm{tot}} \gg N$.

As we will see, the protocol in Theorem 3.1 can be easily modified with no extra overhead to simulate $\mathrm{CZ}_{0,-1}^{\theta} := \mathrm{e}^{\mathrm{i}\theta|11\rangle\langle 11|}$ with any controlled-rotation angle $\theta$ (CZ corresponds to $\theta = \pi$). More generally, the same speed up applies to any two-qubit gate between $0$ and $-1$ by decomposing it into CZ and free single-qubit rotations, which compose a universal gate set.

We present the proof of Theorem 3.1 in Section 3.4. Before that, we develop the idea and introduce several propositions. The technical proofs of the propositions are contained in Section 4.

### 3.1. Generalized GHZ encoding from bosonic squeezing and controlled displacement

At a high level, we follow the idea of [4] and aim to simulate

$$U_{\mathrm{CZ}}^{\mathrm{b}} = (U_{\mathrm{GHZ}}^{\mathrm{b}})^{\dagger} U_{\mathrm{DCZ}}^{\mathrm{b}} U_{\mathrm{GHZ}}^{\mathrm{b}}, \tag{3.5}$$

where we first encode the control qubit $0$ into some "GHZ subspace" of the $N$ ancilla qubits using $U_{\mathrm{GHZ}}^{\mathrm{b}}$, let the ancilla ensemble interact with the target qubit $-1$ using $U_{\mathrm{DCZ}}^{\mathrm{b}}$, and finally reverse the GHZ encoding step. The superscript b means the protocol works in the enlarged boson Hilbert space. See Fig. 2 for an illustration.

We first explain the GHZ encoding step. We demand a generalized version of GHZ encoding:

$$U_{\mathrm{GHZ}}^{\mathrm{b}} \left[ (\eta |0\rangle + \xi |1\rangle)_0 \otimes |\mathbf{0}\rangle \right] = \eta |0\rangle_0 \otimes |-\zeta, \alpha\rangle + \xi |1\rangle_0 \otimes |-\zeta, -\alpha\rangle, \tag{3.6}$$

where

$$\alpha = N^{\frac{1}{2} - \delta_{\alpha}}, \quad \mathrm{e}^{2\zeta} = N^{1 - 2\delta_{\alpha} - 2\delta_{\zeta}}, \tag{3.7}$$

with some constant $0 < \delta_{\alpha} < \delta_{\zeta} < 1/4$. In other words, instead of antipodal points $|\mathbf{0}\rangle, |\mathbf{1}\rangle$ in the DM, we relatively rotate DM to two bosonic coherent states $|\pm\alpha, \zeta\rangle$, depending on the state of control qubit $0$. As mentioned before this state lives in an enlarged bosonic Hilbert space, and we will ultimately implement a qubit Hamiltonian (1.1) that simulates this state. $U_{\mathrm{GHZ}}^{\mathrm{b}}$ is close to the original GHZ encoding [4] at $\alpha = \sqrt{N/2}$, but here we will choose $\alpha \ll \sqrt{N}$ (3.7) in order to suppress simulation errors to $1/\mathrm{poly}(N)$, as we will see. We have also chosen $\delta_{\zeta} > 0$ so that $|-\zeta, \pm\alpha\rangle$ are well separated from each other in phase space, relative to their uncertainty in phase space. The reason for $\delta_{\zeta} > \delta_{\alpha}$ is more subtle, which will become clear in the proof of Theorem 3.1.

The generalized GHZ encoding contains the following subroutines

$$U_{\mathrm{GHZ}}^{\mathrm{b}} = (U_{\mathrm{S}}^{\mathrm{b}})^{\dagger 2} U_{\mathrm{CD}}^{\mathrm{b}} U_{\mathrm{S}}^{\mathrm{b}}, \tag{3.8}$$

where

$$U_S^b = e^{-iT_S^b H_S^b}, \quad U_{CD}^b = e^{-iT_{CD}^b H_{CD}^b} \tag{3.9}$$

stands for squeezing (S) and controlled displacement (CD) respectively. The bosonic squeezing Hamiltonian

$$H_S^b := i\frac{N}{2}\left(B^2 - B^{\dagger 2}\right), \tag{3.10}$$

resembles the usual two-axis-twisting squeezing Hamiltonian $XY+YX$, as used in the previous GHZ encoding protocol [4].

Similarly, the controlled displacement

$$H_{CD}^b = \sqrt{2N}Z_0 \otimes \hat{p} = -i\sqrt{N}Z_0 \otimes \left(B - B^{\dagger}\right), \tag{3.11}$$

which is the bosonic version of controlled rotation $Z_0 \otimes Y$.

During the evolution of $U_{GHZ}$, the boson remains in a squeezed coherent state $|\zeta, \alpha\rangle$. From the exact solution, in order to get (3.6) we require

$$T_S^b = \frac{\zeta}{N} = \left(\frac{1}{2} - \delta_\alpha - \delta_\zeta\right)\frac{\log N}{N}, \tag{3.12}$$

and

$$\alpha = \sqrt{N}T_{CD}^b e^{2NT_S^b}, \quad \Rightarrow \quad T_{CD}^b = N^{-1+\delta_\alpha+2\delta_\zeta}. \tag{3.13}$$

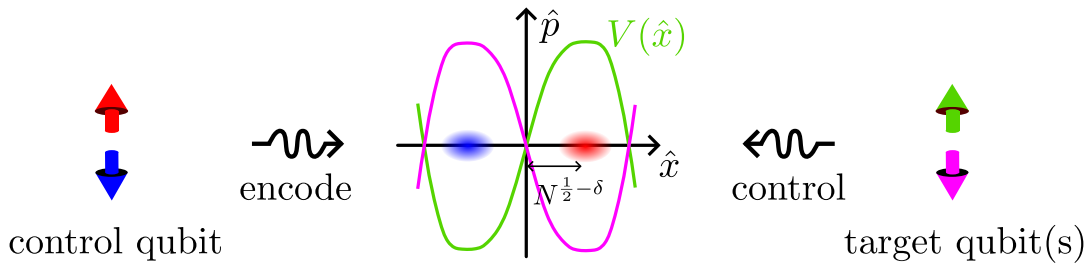

FIG. 2. High-level sketch of the protocol (3.5) in Theorem 3.1. We first perform "GHZ encoding" where the boson is displaced to coherent states at $\hat{x} = \pm N^{\frac{1}{2}-\delta}$ depending on the control qubit $Z_0 = \pm 1$. This encoding step is achieved by squeezing the boson, controlled-displacing the squeezed state, and then unsqueezing while amplifying the signal (3.8). We then use the displacement as an enhanced signal to control the phase of the target qubit $Z_{-1}$. The idea of the latter displacement-controlled-Z is to engineer a potential $V(\hat{x})$ of the boson depending on $Z_{-1}$, which is flat at the two coherent states and almost stabilizes them. Technically we directly engineer a potential as a function of spin polarization $X$. The protocol generalizes to unbounded fan-out gates in Corollary 5.2, because multiple target qubits can be controlled in parallel.

## 3.2. Simulating bosons by qubits

The target Hamiltonians (3.10) and (3.11) are defined by the boson operators. We show they can be effectively simulated by the qubit Hamiltonian (1.1) by the following simulation sequence:

$$\exp\left(-iT_{gad}H_{gad}\right) \approx \exp\left(-iT_{gad}^b H_{gad}^{tr}\right) \approx \exp\left(-iT_{gad}^b H_{gad}^b\right), \tag{3.14}$$

which holds if the boson number is not too large. Here gad = S, CD, and $H_{gad}^{tr}$ is the truncated version of the gadget $H_{gad}^b$, i.e. replacing all $B$ in its expression with $b$. When further expressed in spin operators $X, Y, Z$, $H_{gad}^{tr}$ involves square roots from (2.2), so we expand it to a polynomial of spin operators that can be implemented by the physical $H_{gad}$. More precisely, define

$$q(Z) := \sum_{m=0}^{K-2}\binom{2m}{m}\left(\frac{N-Z}{8N}\right)^m, \tag{3.15}$$

as the $(K-2)$th-order Taylor expansion of $q_0(Z) := \left(1 - \frac{N-Z}{2N}\right)^{-1/2}$, at small $N-Z$. The expansion error at bounded boson number is bounded by the remainder of Taylor expansion:

$$\|(q(Z) - q_0(Z)) P_{\leq n}\| \leq \binom{2(K-1)}{K-1} \left(\frac{n}{4N}\right)^{K-1} \leq \left(\frac{n}{N}\right)^{K-1}. \tag{3.16}$$

We then define

$$H_{\mathrm{CD}} = h_{\mathrm{CD}} \frac{-\mathrm{i}}{2} Z_0 \otimes q(Z) X^+ + \mathrm{H.c.}, \tag{3.17}$$

that approximates (3.11). Here we use a normalization factor $h$ for the Hamiltonian to satisfy (1.2); we will show that it is a constant prefactor.

For S, we use $b^2 = \frac{1}{4N} q_0(Z) X^+ q_0(Z) X^+ + \mathrm{H.c.} = \frac{1}{4N} q_0'(Z)(X^+)^2 + \mathrm{H.c.}$ where

$$\begin{aligned}
q_0'(Z) &= q_0(Z) q_0(Z-1) = p_0^2(Z) \left(1 - \frac{1}{N+Z}\right)^{-1/2} = p_0^2(Z) \left(1 - \frac{p_0^2(Z)}{2N}\right)^{-1/2} \\
&= p_0^2(Z) \sum_{k=0}^{\infty} \binom{2k}{k} \frac{p_0^{2k}(Z)}{(8N)^k} = \sum_{k=0}^{\infty} \binom{2k}{k} (8N)^{-k} \left(1 - \frac{N-Z}{2N}\right)^{-(k+1)} \\
&= \sum_{k=0}^{\infty} \binom{2k}{k} (8N)^{-k} \sum_{m=0}^{\infty} \frac{(k+m)!}{k!m!} \left(\frac{N-Z}{2N}\right)^m \\
&= \sum_{m=0}^{\infty} C_m' \left(\frac{N-Z}{2N}\right)^m,
\end{aligned} \tag{3.18}$$

where

$$C_m' = \sum_{k=0}^{\infty} \frac{(2k)!(k+m)!}{(k!)^3 m!} (8N)^{-k} = 1 + \mathrm{O}\left(\frac{m}{N}\right). \tag{3.19}$$

Here in the second line of (3.18) we have assumed $N - Z \ll N$ to expand in $p_0^2(Z)/N \ll 1$. We define

$$q'(Z) := \sum_{m=0}^{K-2} C_m' \left(\frac{N-Z}{2N}\right)^m, \tag{3.20}$$

so that

$$\|[q'(Z) - q_0'(Z)] P_{\leq n}\| \leq |C_{K-1}'| \left(\frac{n}{4N}\right)^{K-1} \leq 2 \left(\frac{n}{4N}\right)^{K-1}, \tag{3.21}$$

where we have used

$$0 < C_m' < 2, \tag{3.22}$$

from (3.19).

We then define

$$H_{\mathrm{S}} = h_{\mathrm{S}} \frac{\mathrm{i}}{8} q'(Z)(X^+)^2 + \mathrm{H.c.}, \tag{3.23}$$

to simulate $H_{\mathrm{S}}^{\mathrm{b}}$, with a normalization factor $h_{\mathrm{S}}$.

### 3.3. Displacement-controlled-Z

The displacement-controlled-Z (DCZ) unitary in (3.5)

$$U_{\mathrm{DCZ}}^{\mathrm{b}} = \mathrm{e}^{-\mathrm{i} T_{\mathrm{DCZ}}^{\mathrm{b}} H_{\mathrm{DCZ}}^{\mathrm{b}}}, \tag{3.24}$$

is generated by

$$H_{\mathrm{DCZ}}^{\mathrm{b}} = Z_{-1} \otimes L^{\mathrm{b}}, \quad \text{where} \quad L^{\mathrm{b}} \left| -\zeta, \pm\alpha \right\rangle = \pm\overline{X} \left| -\zeta, \pm\alpha \right\rangle, \quad \overline{X} := 2\sqrt{N - \alpha^2}\,\alpha \geq N^{1-\delta_\alpha}. \tag{3.25}$$

$H_{\mathrm{DCZ}}^{\mathrm{b}}$ can be viewed as a $ZZ$-type interaction between qubit $-1$ and the logical-$Z$ operator in the subspace spanned by $\left| -\zeta, \pm\alpha \right\rangle$. Note that $L^{\mathrm{b}}$ is not Hermitian because there is a tiny overlap between the two states — this is not a problem because we will implement it approximately by Hermitian Hamiltonians. Although (3.25) is not a simple Hamiltonian expressed in $B, B^\dagger$, we will demonstrate how to approximate it by qubit Hamiltonians.

One can verify using (3.6) and (3.25) that (3.5) achieves $\mathrm{CZ}_{0,-1}$ for

$$T_{\mathrm{DCZ}}^{\mathrm{b}} = \frac{\pi}{4}\overline{X}^{-1} \leq N^{-1+\delta_\alpha}, \tag{3.26}$$

up to free single-qubit rotations on 0 and $-1$.

To simulate DCZ, we define

$$H_{\mathrm{DCZ}} = h_{\mathrm{DCZ}} Z_{-1} \otimes L, \quad \text{where} \quad L = \overline{X} \sum_{m=0}^{M} C_m \left( \frac{X}{\overline{X}} \right)^{2m+1}. \tag{3.27}$$

**Proposition 3.2.** *Fix any integer $M > 0$. There exist coefficients $\{C_m\}$ satisfying*

$$|C_m| \leq c_{\mathrm{M}} := [M(2M+1)!]^M, \tag{3.28}$$

*such that the corresponding $L$ in (3.27) satisfies*

$$\left\| L \left| -\zeta, \pm\alpha \right\rangle - \left( \pm\overline{X} \left| -\zeta, \pm\alpha \right\rangle \right) \right\| \leq N^{1-\frac{1}{2}\delta_\zeta(M+1)}. \tag{3.29}$$

The error is polynomially small in $N$ for sufficiently large constant $M$ compared to $1/\delta_\zeta$.

The normalization factors introduced above are all mild:

**Proposition 3.3.** *The physical Hamiltonians $H_{\mathrm{gad}}$ with $\mathrm{gad} = \mathrm{CD}, \mathrm{S}, \mathrm{DCZ}$ satisfy normalization (1.2) for*

$$h_{\mathrm{S}} = h_{\mathrm{CD}} = (c_{\mathrm{tot}} K)^{-K}, \tag{3.30}$$

*and*

$$h_{\mathrm{DCZ}} = \frac{1}{2} c_{\mathrm{M}}^{-1} c_{\mathrm{tot}}^{-2M} N^{-2M\delta_\alpha}. \tag{3.31}$$

As a result,

$$T_{\mathrm{gad}} = h_{\mathrm{gad}}^{-1} T_{\mathrm{gad}}^{\mathrm{b}}, \tag{3.32}$$

has little overhead comparing to $T_{\mathrm{gad}}^{\mathrm{b}}$. In particular, we will choose $M\delta_\alpha \ll 1$ for $h_{\mathrm{DCZ}}^{-1}$ to not blow up.

### 3.4. Proof of the $1/N$ time protocol

*Proof of Theorem 3.1.* We rewrite (3.5) using (3.8) and (3.9) as

$$U_{\mathrm{CZ}}^{\mathrm{b}} = \mathcal{T} \mathrm{e}^{-\mathrm{i} \int_0^{T^{\mathrm{b}}} \mathrm{d}t^{\mathrm{b}} H^{\mathrm{b}}(t^{\mathrm{b}})}, \tag{3.33}$$

where $H^{\mathrm{b}}(t^{\mathrm{b}})$ is one of the three $H_{\mathrm{gad}}^{\mathrm{b}}$ (or with a minus sign) depending on where $t^{\mathrm{b}}$ is in the whole protocol. The whole duration

$$T^{\mathrm{b}} = T_{\mathrm{DCZ}}^{\mathrm{b}} + 2T_{\mathrm{CD}}^{\mathrm{b}} + 6T_{\mathrm{S}}^{\mathrm{b}} \leq 3N^{-1+\delta_\alpha+2\delta_\zeta} \leq 3N^{-1/4}, \tag{3.34}$$

is dominated by $T_{\mathrm{CD}}^{\mathrm{b}}$ using (3.13), (3.12) and (3.26). The physical time duration is

$$T = h_{\mathrm{DCZ}}^{-1} T_{\mathrm{DCZ}}^{\mathrm{b}} + 2h_{\mathrm{CD}}^{-1} T_{\mathrm{CD}}^{\mathrm{b}} + 6h_{\mathrm{S}}^{-1} T_{\mathrm{S}}^{\mathrm{b}} \leq 2c_{\mathrm{M}} c_{\mathrm{tot}}^{2M} N^{-1+(2M+1)\delta_\alpha} + 3(c_{\mathrm{tot}} K)^K N^{-1+\delta_\alpha+2\delta_\zeta} \leq N^{-1+\delta_{\mathrm{T}}}, \tag{3.35}$$

using (3.30) and (3.31), where we set

$$\delta_\zeta = M\delta_\alpha, \quad \delta_\alpha = \frac{1}{2M+2}\delta_{\mathrm{T}}. \tag{3.36}$$

This proves (3.3). We can rescale the time $t \to t^{\mathrm{b}}$ so that the true evolution also has the form

$$\mathcal{T}\mathrm{e}^{-\mathrm{i}\int_0^T dt H(t)} =: U_{\mathrm{CZ}} = \mathcal{T}\mathrm{e}^{-\mathrm{i}\int_0^{T^{\mathrm{b}}} dt^{\mathrm{b}} H(t^{\mathrm{b}})}, \tag{3.37}$$

where $H(t^{\mathrm{b}})$ is one of the three $H_{\mathrm{gad}}/h_{\mathrm{gad}}$.

To prove (3.4), observe that both $H$ and $H^{\mathrm{b}}$ are diagonal in the computational basis $|z\rangle_{0,-1}$ of the two data qubits, so

$$\|(U_{\mathrm{CZ}} - \mathrm{CZ}_{0,-1})|\psi\rangle \otimes |\mathbf{0}\rangle\| \le \max_z \left\|\left(U_{\mathrm{CZ}} - U_{\mathrm{CZ}}^{\mathrm{b}}\right)|z\rangle_{0,-1} \otimes |\mathbf{0}\rangle\right\|, \tag{3.38}$$

where we have ignored the data qubits not acted on by the protocol.

According to Duhamel's identity

$$\mathcal{T}\mathrm{e}^{-\mathrm{i}\int_0^T dt H_1(t)} - \mathcal{T}\mathrm{e}^{-\mathrm{i}\int_0^T dt H_2(t)} = -\mathrm{i}\int_0^T dt \mathcal{T}\mathrm{e}^{-\mathrm{i}\int_t^T ds H_1(s)}(H_1(t) - H_2(t))\mathrm{e}^{-\mathrm{i}\int_0^t ds H_2(s)}, \tag{3.39}$$

the simulation error is bounded by

$$\left\|\left(U_{\mathrm{CZ}} - U_{\mathrm{CZ}}^{\mathrm{b}}\right)|z\rangle_{0,-1} \otimes |\mathbf{0}\rangle\right\| = \left\|\left(\mathcal{T}\mathrm{e}^{-\mathrm{i}\int_0^{T^{\mathrm{b}}} dt^{\mathrm{b}} H(t^{\mathrm{b}})} - \mathcal{T}\mathrm{e}^{-\mathrm{i}\int_0^{T^{\mathrm{b}}} dt^{\mathrm{b}} H^{\mathrm{b}}(t^{\mathrm{b}})}\right)|z\rangle_{0,-1} \otimes |\mathbf{0}\rangle\right\|$$
$$\le T^{\mathrm{b}} \max_{t^{\mathrm{b}} \in [0,T^{\mathrm{b}}]} \left\|\left(H(t^{\mathrm{b}}) - H^{\mathrm{b}}(t^{\mathrm{b}})\right)|z\rangle_{0,-1} \otimes |\zeta_{t^{\mathrm{b}}}, \alpha_{t^{\mathrm{b}}}\rangle\right\|, \tag{3.40}$$

Here we have used triangle inequality when taking norm of (3.39), and the fact that the boson mode remains in a coherent state under $H^{\mathrm{b}}(t^{\mathrm{b}})$:

$$|z\rangle_{0,-1} \otimes |\zeta_{t^{\mathrm{b}}}, \alpha_{t^{\mathrm{b}}}\rangle := \mathcal{T}\mathrm{e}^{-\mathrm{i}\int_0^{t^{\mathrm{b}}} ds H^{\mathrm{b}}(s)}|z\rangle_{0,-1} \otimes |\mathbf{0}\rangle. \tag{3.41}$$

Note that $\zeta_{t^{\mathrm{b}}}, \alpha_{t^{\mathrm{b}}}$ depend implicitly on $z$. We then discuss the three cases of $t^{\mathrm{b}}$ in (3.40).

During DCZ gadget, (3.29) yields

$$\left\|\left(h_{\mathrm{DCZ}}^{-1} H_{\mathrm{DCZ}} - H_{\mathrm{DCZ}}^{\mathrm{b}}\right)|z\rangle_{0,-1} \otimes |\zeta_{t^{\mathrm{b}}}, \alpha_{t^{\mathrm{b}}}\rangle\right\| \le N^{1-\frac{1}{2}\delta_\zeta(M+1)} \le N^{1-\frac{1}{2}\delta_\alpha M(M+1)}, \tag{3.42}$$

using (3.36).

For the other gadgets, observe that for any $t^{\mathrm{b}}$ we have bounded displacement

$$|\alpha_{t^{\mathrm{b}}}| \le \alpha, \tag{3.43}$$

and squeezing

$$\mathrm{e}^{-2|\zeta_{t^{\mathrm{b}}}|} \ge \mathrm{e}^{-2\zeta} = N^{-(1-2\delta_\alpha-2\delta_\zeta)}, \tag{3.44}$$

from the solution of the dynamics. The state is therefore always faraway from the cutoff $\hat{n} \approx N$. Let

$$n = 5\alpha^2 = 5N^{1-2\delta_\alpha}, \tag{3.45}$$

we see that $\zeta_{t^{\mathrm{b}}}, \alpha_{t^{\mathrm{b}}}$ always satisfy (2.25) with $\delta = 2\delta_\zeta$. We thus apply Lemma 2.3 and get (ignoring the subscripts $t^{\mathrm{b}}$ for notational simplicity)

$$\left\|h_{\mathrm{gad}}^{-1} H_{\mathrm{gad}} P_{>n} |z\rangle_{0,-1} \otimes |\zeta, \alpha\rangle\right\|, \quad \left\|H_{\mathrm{gad}}^{\mathrm{b}} P_{>n} |z\rangle_{0,-1} \otimes |\zeta, \alpha\rangle\right\| \le \mathrm{e}^{-N^{2\delta_\zeta(1-2\delta_\alpha)}/10}, \tag{3.46}$$

where we have used (2.23) for $\mathcal{O} = H_{\mathrm{gad}}, H_{\mathrm{gad}}^{\mathrm{b}}$ with a constant $d$ and a $c_d$ that is at most $\mathrm{poly}(N)$. In (3.46) we have adjusted the exponential to absorb the polynomial prefactors including $c_d$.

For gad = S, CD we have

$$
\begin{aligned}
&\left\| \left( h_{\mathrm{gad}}^{-1} H_{\mathrm{gad}} - H_{\mathrm{gad}}^{\mathrm{b}} \right) |z\rangle_{0,-1} \otimes |\zeta, \alpha\rangle \right\| \\
&\leq \left\| \left( h_{\mathrm{gad}}^{-1} H_{\mathrm{gad}} - H_{\mathrm{gad}}^{\mathrm{b}} \right) |z\rangle_{0,-1} \otimes P_{\leq n} |\zeta, \alpha\rangle \right\| + \left\| \left( h_{\mathrm{gad}}^{-1} H_{\mathrm{gad}} - H_{\mathrm{gad}}^{\mathrm{b}} \right) |z\rangle_{0,-1} \otimes P_{>n} |\zeta, \alpha\rangle \right\| \\
&\leq \left\| \left( h_{\mathrm{gad}}^{-1} H_{\mathrm{gad}} - H_{\mathrm{gad}}^{\mathrm{tr}} \right) P_{\leq n} \right\| + 2 \mathrm{e}^{-N^{2\delta_\zeta (1 - 2\delta_\alpha)}/10} \\
&\leq 2^K N^{2 - 2\delta_\alpha (K-1)}.
\end{aligned}
\tag{3.47}
$$

Here we have used triangle inequality and (3.46) to get the third line, where the first term is also changed from $H_{\mathrm{gad}}^{\mathrm{b}}$ to $H_{\mathrm{gad}}^{\mathrm{tr}}$ because of the truncation $P_{\leq n}$. The last line of (3.47) combines

$$
\begin{aligned}
\left\| (h_{\mathrm{CD}}^{-1} H_{\mathrm{CD}} - H_{\mathrm{CD}}^{\mathrm{tr}}) P_{\leq n} \right\| &= \frac{1}{2} \left\| \left[ Z_0 \otimes (q(Z) - q_0(Z)) X^+ + \mathrm{H.c.} \right] P_{\leq n} \right\| \\
&\leq 2N \left\| (q(Z) - q_0(Z)) P_{\leq n+1} \right\| \leq 2N (6N^{-2\delta_\alpha})^{K-1},
\end{aligned}
\tag{3.48}
$$

from (3.16), and

$$
\begin{aligned}
\left\| (h_{\mathrm{S}}^{-1} H_{\mathrm{S}} - H_{\mathrm{S}}^{\mathrm{tr}}) P_{\leq n} \right\| &= \frac{1}{8} \left\| \left[ \mathrm{i} \left( q'(Z) - q_0'(Z) \right) (X^+)^2 + \mathrm{H.c.} \right] P_{\leq n} \right\| \\
&\leq N^2 \left\| (q'(Z) - q_0'(Z)) P_{\leq n+2} \right\| \leq 2N^2 (6N^{-2\delta_\alpha}/4)^{K-1},
\end{aligned}
\tag{3.49}
$$

from (3.21), where the stretched exponential term $\mathrm{e}^{-N^\delta/10}$ is absorbed.

We set

$$
M = \left\lfloor 2\sqrt{K} \right\rfloor + 1 \geq 2\sqrt{K},
\tag{3.50}
$$

where $\lfloor \cdot \rfloor$ is the floor function, so that the bound (3.47) is larger than (3.42). Putting (3.36), (3.34), (3.47) into (3.40) and (3.38),

$$
\left\| \langle \mathbf{0} | \left( U_{\mathrm{CZ}} - \mathrm{CZ}_{0,-1} \right) | \mathbf{0} \rangle \right\| \leq N^{2 - 2\delta_\alpha (K-1)} = N^{2 - \delta_{\mathrm{T}} \frac{K-1}{M+1}} \leq N^{2 - \delta_{\mathrm{T}} (\sqrt{K} - 1)/2},
\tag{3.51}
$$

using $M \leq 2\sqrt{K} + 1$. This finishes the proof. $\qquad\square$

## 4. TECHNICAL INGREDIENTS FOR THE $1/N$ TIME PROTOCOL

### 4.1. Logical operator for two coherent states

*Proof of Proposition 3.2.* Because $L$ is an odd function of $X$, we can focus on the $+$ case of (3.29) due to $X \leftrightarrow -X$ symmetry.

We write $L = \overline{X} f(x)$ where

$$
x := X / \overline{X}
\tag{4.1}
$$

and

$$
f(x) = \sum_{m=0}^{M} C_m x^{2m+1}.
\tag{4.2}
$$

We choose $C_m$ so that

$$
f(1) = 1, \quad \text{and} \quad f^{(k)}(1) = 0, \quad \forall k = 1, 2, \cdots, M.
\tag{4.3}
$$

In other words, $f(x)$ is flat at $x = 1$ up to $M$-th order in Taylor expansion. Plugging (4.2) in (4.3) yields

$$
\sum_m A_{km} C_m = v_k,
\tag{4.4}
$$

where $v_k = \delta_{k0}$, and $A_{km} = (2m+1)_k$ is a $(M+1)$-dimensional square matrix. Here $(m)_k := m(m-1)\cdots(m-k+1)$. Observe that $(2m+1)_k = \sum_{j=0}^{k} S_{kj}(2m+1)^j$ where matrix $S$ is a triangular matrix with $\mathrm{Det}(S) = 1$. So

$$\mathrm{Det}(A) = \mathrm{Det}\left(S\widetilde{A}\right) = \mathrm{Det}\left(\widetilde{A}\right) = \prod_{0 \le m < m' \le M} 2(m'-m) = 2^{M(M+1)/2} \prod_{m=1}^{M} m! \ge 1, \tag{4.5}$$

where $\widetilde{A}_{jm} = (2m+1)^j$ is a Vandermonde matrix with known determinant. In particular, $\mathrm{Det}(A) > 0$ so (4.4) has a unique solution for $C_m$.

Using $C = A^{-1}v$ and the inversion matrix formula, we get (3.28):

$$|C_m| = \frac{\mathrm{Det}(\overline{A}_{0m})}{\mathrm{Det}(A)} \le \frac{1}{\mathrm{Det}(A)} M^M \left[(2M+1)!\right]^M \le \left[M(2M+1)!\right]^M, \tag{4.6}$$

where $\overline{A}_{0m}$ is the matrix $A$ with row 0 and column $m$ deleted, whose determinant is bounded by its matrix elements $|A_{km}| \le (2M+1)!$.

We then show that $|-\zeta, \alpha\rangle$ is mainly supported in a narrow range of $|x-1| \ll 1$, so that $L$ acts almost as a constant. To this end, we first bound the range of boson number using Lemma 2.2 where (2.14) is satisfied due to (3.7):

$$\|(I-\Pi)|-\zeta, \alpha\rangle\| \le 2\exp\left[1 - \frac{0.6}{e^\zeta \alpha}\Delta\right] \le 2\exp\left[1 - 0.6 N^{2\delta_\alpha}\right], \tag{4.7}$$

where $\Pi$ projects onto the subspace of $-\Delta \le \hat{n} - \alpha^2 \le \Delta$ with

$$\Delta := N e^\zeta / \alpha = N^{1-\delta_\zeta}. \tag{4.8}$$

We then focus on subspace $\Pi$, where we show concentration of $X$:

**Lemma 4.1.** *For any even constant $\ell > 0$,*

$$\left\| P_{|X-\overline{X}| > \Delta_X} \Pi |-\zeta, \alpha\rangle \right\| \le c_{\mathrm{L}} N^{-\frac{1}{4}\delta_\zeta \ell}, \tag{4.9}$$

*where constant $c_{\mathrm{L}}$ is determined by $\ell$, and*

$$\Delta_X = N^{1 - \delta_\alpha - \frac{1}{2}\delta_\zeta}. \tag{4.10}$$

As a result,

$$\begin{aligned}
\left\| \left(I - P_{|X-\overline{X}| \le \Delta_X} \Pi\right) |-\zeta, \alpha\rangle \right\| &= \left\| \left(I - \Pi + P_{|X-\overline{X}| > \Delta_X} \Pi\right) |-\zeta, \alpha\rangle \right\| \\
&\le 2\exp\left[1 - 0.6 N^{2\delta_\alpha}\right] + c_{\mathrm{L}} N^{-\frac{1}{4}\delta_\zeta \ell} \le 2 c_{\mathrm{L}} N^{-\frac{1}{4}\delta_\zeta \ell},
\end{aligned} \tag{4.11}$$

where we have used triangle inequality, (4.7) and (4.9). Finally,

$$\begin{aligned}
\left\| \left(L - \overline{X}\right) |-\zeta, \alpha\rangle \right\| &\le \left\| \left(L - \overline{X}\right) P_{|X-\overline{X}| \le \Delta_X} \Pi |-\zeta, \alpha\rangle \right\| + \|L - \overline{X}\| \left\| \left(I - P_{|X-\overline{X}| \le \Delta_X} \Pi\right) |-\zeta, \alpha\rangle \right\| \\
&\le \left\| \left(L - \overline{X}\right) P_{|X-\overline{X}| \le \Delta_X} \right\| + \left(\|L\| + \overline{X}\right) 2 c_{\mathrm{L}} N^{-\frac{1}{4}\delta_\zeta \ell} \\
&\le \overline{X} \max_{|x-1| \le \Delta_X/\overline{X}} |f(x) - 1| + 2 c_{\mathrm{L}} N^{-\frac{1}{4}\delta_\zeta \ell} \overline{X} \left[1 + \sum_{m=0}^{M} |C_m| \left(\frac{N}{\overline{X}}\right)^{2m+1}\right] \\
&\le \overline{X} \frac{1}{(M+1)!} \left|f^{(M+1)}(1)\right| N^{-\frac{1}{2}\delta_\zeta (M+1)} + 3 c_{\mathrm{L}} |C_M| \overline{X} N^{\delta_\alpha (2M+1) - \frac{1}{4}\delta_\zeta \ell} \\
&\le N^{1 - \frac{1}{2}\delta_\zeta (M+1)}.
\end{aligned} \tag{4.12}$$

Here the first two lines come from triangle inequality and (4.11). The third line uses $\|X\| = N$. In the fourth line, we have used Taylor expansion's remainder, $\overline{X} \ge N^{1-\delta_\alpha}$ from (3.25) and $\Delta_X/\overline{X} \le N^{-\frac{1}{2}\delta_\zeta}$ from (4.10). To get the last line, we set

$$\ell \ge 2M + 2 + 4(2M+1)\delta_\alpha/\delta_\zeta, \tag{4.13}$$

so that the first term dominates, and substitute $\overline{X} \to N$ to absorb the constant prefactors. $\qquad \square$

## 4.2. Concentration in $X$ for coherent states

*Proof of Lemma 4.1.* We express the spin operator by bosons:

$$P(X - \overline{X})P = P\sqrt{N - \alpha^2}\left[(B - \alpha) + (B^\dagger - \alpha) + E(f_0; g_0)\right]P, \tag{4.14}$$

where

$$E(f; g) := f(\hat{n}) + [g(\hat{n})B + \text{H.c.}], \tag{4.15}$$

and the particular functions in (4.14) are

$$f_0(\hat{n}) = 0, \quad g_0(\hat{n}) = \left(\sqrt{\frac{N - \hat{n}}{N - \alpha^2}} - 1\right). \tag{4.16}$$

Let $\Pi'$ be the projector onto subspace $-3\Delta \leq \hat{n} - \alpha^2 \leq 3\Delta$, i.e. a slightly wider window than $\Pi$. Since $g_0(\hat{n})$ is a smooth function of $\hat{n}$ that changes in scale $\sim N$, for any constant $m \leq \ell^2$ with $\ell$ indepedent of $N$ we have

$$\|g_0(\hat{n})\Pi'\| \leq c_{0,\ell}\frac{\Delta}{N}, \quad \left\|g_0^{(m)}(\hat{n})\Pi'\right\| \leq c_{0,\ell}N^{-m} \quad (m \geq 1), \tag{4.17}$$

for the discrete derivatives defined recursively by

$$g_0^{(m+1)}(\hat{n}) := g_0^{(m)}(\hat{n}) - g_0^{(m)}(\hat{n}), \tag{4.18}$$

with $g_0^{(0)}(\hat{n}) = g_0(\hat{n})$. $c_{0,\ell}$ is a constant determined by $\ell$.

Aiming for a final Markov's inequality for (4.9), consider

$$\left\langle \Pi\left(X - \overline{X}\right)^\ell \Pi \right\rangle_{-\zeta, \alpha} = \left\langle \Pi P\left(X - \overline{X}\right)^\ell P\Pi \right\rangle_{-\zeta, \alpha} = \left\langle \Pi\left[P(X - \overline{X})P\right]^\ell \Pi \right\rangle_{-\zeta, \alpha} = (N - \alpha^2)^{\frac{\ell}{2}}\sum_\mu \mathbb{X}_\mu, \tag{4.19}$$

with an even constant $\ell > 0$. Here each term is of the form

$$\mathbb{X}_\mu = \langle \Pi \mathbb{Y}_1 \cdots \mathbb{Y}_\ell \Pi \rangle_{-\zeta, \alpha}, \tag{4.20}$$

where each $\mathbb{Y}_m$ is one of the three choices in (4.14), and there are $3^\ell$ number of terms in (4.19). For each term $\mu$, we use (2.5) to commute $(B - \alpha), (B^\dagger - \alpha)$ through $E(f; g)$ to the right (we do not commute $(B - \alpha), (B^\dagger - \alpha)$ through each other). More precisely, for each commutation we have

$$\begin{aligned}(B - \alpha)E(f; g) &= f(\hat{n} + 1)B + [g(\hat{n} + 1)B + \text{H.c.}]B + g(\hat{n}) - \alpha\{f(\hat{n}) + [g(\hat{n})B + \text{H.c.}]\} \\ &= E(f'; g')(B - \alpha) + E(f''; g'')e^\zeta,\end{aligned} \tag{4.21}$$

where

$$f'(\hat{n}) = f(\hat{n}+1), \quad g'(\hat{n}) = g(\hat{n}+1), \quad f''(\hat{n})e^\zeta = g(\hat{n}) + \alpha\left[f(\hat{n}+1) - f(\hat{n})\right], \quad g''(\hat{n})e^\zeta = \alpha\left[g(\hat{n}+1) - g(\hat{n})\right]. \tag{4.22}$$

The case commuting with $B^\dagger - \alpha$ is similar with substitutions $\hat{n} + 1 \to \hat{n} - 1$.

Starting from the original $f_0, g_0$, after a constant number $\leq \ell^2$ of iterations given by (4.22) or its $\hat{n} - 1$ counterpart, any $f, g$ we get is a sum over $e^{-\zeta m'}\alpha^m g_0^{(m)}(\hat{n} + m'')$ where $0 \leq m \leq m' \leq \ell^2$ and $|m''| \leq \ell^2$. From (4.17) and $e^{-\zeta}\alpha = N/\Delta$, the dominant term is always $m = 0$ so

$$\|f(\hat{n})\Pi''\|, \|g(\hat{n})\Pi''\| \leq \frac{1}{4}c_\ell\Delta/N, \tag{4.23}$$

where $\Pi''$ projects onto subspace $-2\Delta \leq \hat{n} - \alpha^2 \leq 2\Delta$, and $c_\ell$ is a constant determined by $\ell$. As a result,

$$\begin{aligned}\|\Pi''E(f; g)\Pi''\| &\leq \|f(\hat{n})\Pi''\| + \|\Pi''g(\hat{n})B\Pi''\| + \left\|\Pi''B^\dagger g(\hat{n})\Pi''\right\| \\ &\leq \|f(\hat{n})\Pi''\| + 2\|g(\hat{n})\Pi''\|\|\Pi''B\Pi''\| \leq c_\ell\frac{\Delta}{N}\alpha = c_\ell e^\zeta,\end{aligned} \tag{4.24}$$

using $\|\Pi''B\Pi''\| \leq \sqrt{\alpha^2 + 2\Delta + 1}$ and (4.8).

Commuting the operators through in $\mathbb{X}_\mu$, we have

$$\mathbb{X}_\mu = \sum \left\langle \Pi\left(\Pi''E(f_1; g_1)\Pi''\right)\cdots\left(\Pi''E(f_{\ell''}; g_{\ell''})\Pi''\right)\mathrm{e}^{\zeta(\ell-\ell''-\ell')}\mathbb{B}_1\cdots\mathbb{B}_{\ell'}\Pi\right\rangle_{-\zeta,\alpha}. \tag{4.25}$$

There are a constant number of terms in the sum of (4.25) determined by $\ell$, and each $\mathbb{B}_\nu \in \{(B-\alpha), (B^\dagger-\alpha)\}$, with $0 \leq \ell', \ell'' \leq \ell$. All $f_m, g_m$ satisfy (4.23) because they come from finite number of iterations in (4.22). We have also inserted $\Pi''$ in (4.25) because the outer $\Pi$ ensures that the boson number for each individual $E$ is bounded.

From Wick's theorem (2.11),

$$\|\mathbb{B}_1\cdots\mathbb{B}_{\ell'}\,|-\zeta,\alpha\rangle\|^2 = \left\langle \mathbb{B}_{\ell'}^\dagger\cdots\mathbb{B}_1^\dagger\mathbb{B}_1\cdots\mathbb{B}_{\ell'}\right\rangle_{-\zeta,\alpha} \leq (2\ell'-1)!!\mathrm{e}^{2\zeta\ell'}, \tag{4.26}$$

where there are $(2\ell'-1)!!$ different contractions and each pairwise contraction is bounded by $\mathrm{e}^{2\zeta}$ according to (2.12). Therefore,

$$\begin{aligned}
\|\mathbb{B}_1\cdots\mathbb{B}_{\ell'}\Pi\,|-\zeta,\alpha\rangle\| &\leq \|\mathbb{B}_1\cdots\mathbb{B}_{\ell'}\,|-\zeta,\alpha\rangle\| + \|\mathbb{B}_1\cdots\mathbb{B}_{\ell'}P(I-\Pi)\,|-\zeta,\alpha\rangle\| + \|\mathbb{B}_1\cdots\mathbb{B}_{\ell'}(I-P)\,|-\zeta,\alpha\rangle\| \\
&\leq \sqrt{(2\ell'-1)!!}\mathrm{e}^{\zeta\ell'} + \|\mathbb{B}_1\cdots\mathbb{B}_{\ell'}P\|\,\|(I-\Pi)\,|-\zeta,\alpha\rangle\| + 2c_d\mathrm{e}^{-N^{2\delta_\alpha+2\delta_\zeta}/10} \\
&\leq \sqrt{2(2\ell'-1)!!}\mathrm{e}^{\zeta\ell'}, 
\end{aligned} \tag{4.27}$$

where we have used (4.7) and Lemma 2.3 for the second and third term respectively, which are all exponentially small in $N$ and thus subdominant comparing to the first term.

Using (4.23) and (4.27), we have

$$\begin{aligned}
|\mathbb{X}_\mu| &\leq c_\ell'\mathrm{e}^{\zeta\ell''}\mathrm{e}^{\zeta(\ell-\ell''-\ell')}\sqrt{2(2\ell'-1)!!}\mathrm{e}^{\zeta\ell'} \\
&\leq \sqrt{2(2\ell'-1)!!}c_\ell'\mathrm{e}^{\zeta\ell} = \sqrt{2(2\ell'-1)!!}c_\ell'N^{(\frac{1}{2}-\delta_\alpha-2\delta_\zeta)\ell},
\end{aligned} \tag{4.28}$$

where constant $c_\ell'$ is determined by $\ell$. From (4.19) this leads to

$$\left\langle \Pi\left(X-\overline{X}\right)^\ell\Pi\right\rangle_{-\zeta,\alpha} \leq 3^\ell\sqrt{2(2\ell'-1)!!}c_\ell'N^{(1-\delta_\alpha-2\delta_\zeta)\ell}. \tag{4.29}$$

Finally, we apply Markov's inequality:

$$\left\|P_{X-\overline{X}>\Delta_X}\Pi|-\zeta,\alpha\rangle\right\|^2 \leq \Delta_X^{-\ell}\left\langle \Pi\left(X-\overline{X}\right)^\ell\Pi\right\rangle_{-\zeta,\alpha} \leq 3^\ell\sqrt{2(2\ell-1)!!}c_\ell'N^{-\frac{1}{2}\delta_\zeta\ell}, \tag{4.30}$$

which yields (4.9) for $c_\mathrm{L} = \sqrt{3^\ell\sqrt{2(2\ell-1)!!}c_\ell'}$ and (4.10). $\qquad\square$

### 4.3. Normalization

*Proof of Proposition 3.3.* We first deal with the simplest case DCZ.

$$L = \overline{X}\sum_{k=1,3,\cdots,2M+1}\overline{X}^{-k}J_k^L\sum_{1\leq i_1<\cdots<i_k\leq N}X_{i_1}\cdots X_{i_k}, \tag{4.31}$$

where

$$J_k^L = C_{\frac{k-1}{2}} + \mathrm{O}(N\overline{X}^{-2}) = C_{\frac{k-1}{2}} + \mathrm{O}(N^{-1+2\delta_\alpha}), \tag{4.32}$$

using (3.25). The second term $\mathrm{O}(N^{-1+2\delta_\alpha})$ represents contributions from expanding $X^{k'}$ for $k' = k+2, k+4, \cdots$ and getting a term like $X_{i_0}X_{i_0}X_{i_1}\cdots X_{i_k}$ with repeated qubits (collisions). Such contributions are suppressed by a factor $(N\overline{X}^{-2})^{\frac{k'-k}{2}}$ because there is a $\overline{X}^{-k'}$ factor in front of $X^{k'}$, yet there are at most $\mathrm{O}(N^{(k'-k)/2})$ choices of the repeated qubits: For a collision to happen, the first qubit $i_0$ has $N$ choices, but the second qubit is fixed once the first is chosen.

As a result, $H_{\mathrm{DCZ}} = \sum_k N_{\mathrm{tot}}^{2-k} \sum_{1 \le i_1 < \cdots < i_{k-1} \le N} J_{-1, i_1 \cdots i_{k-1}}^{3, 1 \cdots 1} Z_{-1} \otimes X_{i_1} \cdots X_{i_{k-1}}$ where

$$\left| J_{-1, i_1 \cdots i_{k-1}}^{3, 1 \cdots 1} \right| = h_{\mathrm{DCZ}} \left( \frac{\overline{X}}{N_{\mathrm{tot}}} \right)^{2-k} \left[ C_{\frac{k}{2}-1} + \mathrm{O}(N^{-1+2\delta_\alpha}) \right] \le c_{\mathrm{tot}}^{k-2} h_{\mathrm{DCZ}} N^{2M\delta_\alpha} 2 \left[ M(2M+1)! \right]^M \quad (4.33)$$

using (3.1), (3.31) and (3.28). We can thus choose (3.31) to satisfy normalization (1.2) because $k \le 2M+2$.

We then treat S. From (3.20),

$$\begin{aligned}
q'(Z) &= \sum_{m=0}^{K-2} C_m' \sum_{k=0}^m \binom{m}{k} 2^{-(m-k)} \left( \frac{-1}{N} \right)^k Z^k = \sum_{k=0}^{K-2} \left( \frac{-2}{N} \right)^k \left[ \sum_{m=k}^{K-2} C_m' \binom{m}{k} 2^{-m} \right] Z^k \\
&= \sum_{k=0}^{K-2} \left( \frac{-2}{N} \right)^k J_k^{q'} \sum_{1 \le i_1 < \cdots < i_k \le N} Z_{i_1} \cdots Z_{i_k},
\end{aligned} \quad (4.34)$$

where

$$\left| J_k^{q'} \right| = k! \sum_{m=k}^{K-2} C_m' \binom{m}{k} 2^{-m} + \mathrm{O}(N^{-1}) \le k! \left( \sum_{m=k}^{K-2} 2 \right) + 1 \le 2k!K, \quad (4.35)$$

using (3.22) and that $N$ is sufficiently large. Similar to (4.31), the first term in (4.35) comes from expanding $Z^k$ and get $Z_{i_1} \cdots Z_{i_k}$ with $i_1 < \cdots < i_k$, each repeated $k!$ times. We then have

$$H_{\mathrm{S}} = h_{\mathrm{S}} \frac{\mathrm{i}}{4} \sum_{k=0}^{K-2} \left( \frac{-2}{N} \right)^k J_k^{q'} \sum_{1 \le i_1 < \cdots < i_k \le N} Z_{i_1} \cdots Z_{i_k} \sum_{j_1 < j_2} X_{j_1}^+ X_{j_2}^+ + \mathrm{H.c.}. \quad (4.36)$$

When expressed as (1.2),

$$\sum_{a_1 \cdots a_k} \left| J_{i_1 \cdots i_k}^{a_1 \cdots a_k} \right| \le c_{\mathrm{tot}}^{2-k} \left( 2 \times \frac{1}{4} \times 2^2 \right) h_{\mathrm{S}} \binom{k}{2} \left( 2^{k-2} |J_{k-2}^{q'}| + \mathrm{O}(N^{-1}) \right) \le c_{\mathrm{tot}}^{-k} h_{\mathrm{S}} k^2 2^k k! K \le c_{\mathrm{tot}}^{-K} h_{\mathrm{S}} K^K. \quad (4.37)$$

Here $2 \times \frac{1}{4} \times 2^2$ comes from the Hermitian conjugate, the numerical prefactor and $X_i^+ = X_i + \mathrm{i} Y_i$. We have again used the fact that when there are collisions between $i_\mu$ and $j_\nu$ in (4.36), the contribution is suppressed by at least $\mathrm{O}(1/N)$. (4.37) indicates $h_{\mathrm{S}} = (c_{\mathrm{tot}} K)^{-K}$ for normalization (1.2) to hold.

For CD, observe that $q(Z)$ (3.15) when expressed as (3.20) also has coefficients satisfying (3.22), so the corresponding

$$|J_k^q| \le 2k!K, \quad (4.38)$$

is also the same. Similar to (4.37), when expressing $H_{\mathrm{CD}}$ as (1.2) its coefficients satisfy

$$\sum_{a_1 \cdots a_k} \left| J_{i_1 \cdots i_k}^{a_1 \cdots a_k} \right| \le c_{\mathrm{tot}}^{2-k} \left( 2 \times \frac{1}{2} \times 2 \right) h_{\mathrm{CD}}(k-1) \left( 2^{k-2} |J_{k-2}^q| + \mathrm{O}(N^{-1}) \right) \le c_{\mathrm{tot}}^{-k} h_{\mathrm{CD}} k^2 k! K \le c_{\mathrm{tot}}^{-K} h_{\mathrm{CD}} K^K. \quad (4.39)$$

So we can also choose $h_{\mathrm{CD}} = (c_{\mathrm{tot}} K)^{-K}$. This finishes the proof. □

## 5. IMPLICATIONS OF THE $1/N$ TIME PROTOCOL

### 5.1. Trading space with time for any quantum circuit

A direct consequence of Theorem 3.1 is that, by paying a space overhead $N/N_{\mathrm{d}}$, one can speed up any quantum circuit evolution on $N_{\mathrm{d}}$ qubits by roughly the same factor $N/N_{\mathrm{d}}$ using all-to-all Hamiltonians. Intriguingly, this tradeoff approximately preserves the spacetime volume. The formal statement is:

**Corollary 5.1** (Trading space for time)**.** *Consider the same setting as Theorem 3.1. Any quantum circuit $U_{\mathrm{cir}}$ of depth $D$ on the $N_{\mathrm{d}} = N_{\mathrm{tot}} - N$ data qubits can be simulated by $H(t)$ in (1.1) with time*

$$T = N_{\mathrm{d}} N^{-1+\delta_{\mathrm{T}}} D, \quad (5.1)$$

*and polynomially error $N_{\mathrm{d}} D \epsilon$, where $\epsilon$ is given in (3.2).*

*Proof.* We simulate each CZ gate in the circuit by the protocol in Theorem 3.1, while the single-qubit rotations in the circuit can be implemented exactly. The CZ gates that act in parallel in $U_{\text{cir}}$ have to be simulated sequentially, so the protocol is repeated $\leq N_{\text{d}}D$ times, resulting in total evolution time (5.1) from (3.3).

From triangle inequality,

$$
\begin{aligned}
\|(U_2'U_1' - U_2U_1)|\psi\rangle \otimes |\mathbf{0}\rangle\| &= \|[U_2'(U_1' - U_1) + (U_2' - U_2)U_1]|\psi\rangle \otimes |\mathbf{0}\rangle\| \\
&\leq \|U_2'(U_1' - U_1)|\psi\rangle \otimes |\mathbf{0}\rangle\| + \|(U_2' - U_2)U_1|\psi\rangle \otimes |\mathbf{0}\rangle\| \\
&= \|(U_1' - U_1)|\psi\rangle \otimes |\mathbf{0}\rangle\| + \|(U_2' - U_2)U_1|\psi\rangle \otimes |\mathbf{0}\rangle\|,
\end{aligned}
\tag{5.2}
$$

so if $U_1$ acts only on the data qubit,

$$
\max_{\psi} \|(U_2'U_1' - U_2U_1)|\psi\rangle \otimes |\mathbf{0}\rangle\| \leq \max_{\psi} \|(U_1' - U_1)|\psi\rangle \otimes |\mathbf{0}\rangle\| + \max_{\psi} \|(U_2' - U_2)|\psi\rangle \otimes |\mathbf{0}\rangle\|.
\tag{5.3}
$$

This relation can be iterated by setting $U_2 \to U_3U_2$ etc, so that as long as all $U_m$ acts only on the data qubit,

$$
\max_{\psi} \|(U_m' \cdots U_2'U_1' - U_m \cdots U_2U_1)|\psi\rangle \otimes |\mathbf{0}\rangle\| \leq \sum_{m'=1}^{m} \max_{\psi} \|(U_{m'}' - U_{m'})|\psi\rangle \otimes |\mathbf{0}\rangle\|,
\tag{5.4}
$$

i.e. the approximation errors just add up.

Letting each $U_m$ to be one gate in (1.5), the total simulation error of our protocol is the total number of gates $\leq N_{\text{d}}D$ times the individual simulation error $\epsilon$ in (3.2), which yields the advertised error $N_{\text{d}}D\epsilon$. $\qquad\square$

Later on we will also consider quantum circuits equipped with unbounded fan-out gates. Corollary 5.1 generalizes easily to such a larger class of quantum circuits, because any unbounded fan-out gate can be decomposed to a depth-$\log_2(N_{\text{d}})$ usual quantum circuit. This extra logarithmic factor does not change the essence of the space-time tradeoff here. We will consider unbounded fan-out because there are interesting quantum circuits aided by them, which can be simulated by Hamiltonians in a much faster timescale than predicted in Corollary 5.1. In particular, the results below only need a comparable amount of ancilla $N = \Theta(N_{\text{d}})$.

### 5.2. Fast protocols for unbounded fan-out and related circuits

Theorem 3.1 can be generalized to show that the unbounded fan-out gate [10], which is equivalent (up to single-qubit rotations) to $\text{CZ}_{0,S} := \prod_{i \in S} \text{CZ}_{0,i}$ with an arbitrarily large set $S$ of target qubits, can be simulated in roughly $1/N$ time with $1/\text{poly}(N)$ error.

More generally, a depth-$D$ quantum circuit with unbounded fan-out gates can be simulated by all-to-all Hamiltonians in time $T \ll D$, if there are not many entangling gates acting in parallel. More precisely, consider a unitary on data qubits $U_{\text{cir}} = U_{\text{cir},D} \cdots U_{\text{cir},2}U_{\text{cir},1}$ where each layer $U_{\text{cir},m}$ is either single-qubit rotations, or of the form

$$
U_{\text{cir},m} = \prod_{j \in S} \text{CZ}_{j,R(j)}^{\theta(j,R(j))},
\tag{5.5}
$$

where $S$ is a subset of control qubits $j$, and $R(j)$ is the set of target qubits controlled by $j$. $R(j)$ with different $j$ are allowed to overlap. $S, R(j)$ and the corrotation angles could depend on $m$.

**Corollary 5.2** (Fast unbounded fan-out). *Consider the same setting as Theorem 3.1 with $N_{\text{d}} = N_{\text{tot}} - N \leq (c_{\text{tot}} - 1)N$ data qubits. Consider any depth-$D$ quantum circuit $U_{\text{cir}}$ on the data qubits with possibly unbounded fan-out gates, whose entangling-gate layers satisfy (5.5) with $|S| \leq W$. Such a circuit can be simulated by an all-to-all Hamiltonian (1.1) in time*

$$
T \leq WDN^{-1+\delta_{\text{r}}},
\tag{5.6}
$$

*with polynomially small error $WDN_{\text{d}}\epsilon$, where $\epsilon$ given in (3.2).*

*Proof.* In the proof of Corollary 5.1, we have shown that both time and simulation error add up from the simulation of individual gates, so here it suffices to prove Corollary 5.2 for a single unbounded fan-out gate $\text{CZ}_{0,S}$ with $W = D = 1$.

We use the same protocol as Theorem 3.1, with the only difference that during DCZ we couple the ancillae to not just one qubit $-1$ but all qubits $j \in S$. This does not change the Hamiltonian normalization because we are just turning on more couplings, so $T$ is given by (5.6) with $W = D = 1$.

Since the couplings to different $j$ commute, the new protocol is $\prod_{j \in S} U_j$ where each $U_j$ is of the form $U_{\mathrm{CZ}}$ in (3.37) that simulates $\mathrm{CZ}_{0,j}$. The $U_j$s commute with each other, and obey error bound

$$\|(U_j - \mathrm{CZ}_{0,j}) |\psi\rangle \otimes |\mathbf{0}\rangle\| \le \epsilon, \tag{5.7}$$

according to Theorem 3.1. Triangle inequality (5.4) then ensures a total error $|S|\epsilon$, which gives rise to total error $WDN_{\mathrm{d}}\epsilon$. $\qquad\square$

An interesting class of circuits using a small number of unbounded fan-out is the following:

**Example 5.3** (Fast simulation for circuits manipulating the Hamming weight)**.** *Consider the same setting as Corollary 5.2. Consider the following unitary gates on $n + 1 \le N_{\mathrm{d}}$ data qubits that map $|z\rangle |y\rangle \to |z\rangle |y \oplus g(z)\rangle$ where $g(z) = 1$ iff*

$$
\begin{array}{lll}
|z| > 0 : \mathrm{Or}, & |z| = n : \mathrm{And}\ (\mathrm{Toffoli}), & |z| \ge \mathtt{t} : \ \mathrm{Threshold}[\mathtt{t}] \\
|z|\ mod\ \mathtt{q} = 0 : \mathrm{Mod}[\mathtt{q}], & |z| = \mathtt{t} : \mathrm{Exact}[\mathtt{t}],
\end{array} \tag{5.8}
$$

*where $z$ is a bitstring of length $n$, and $|z|$ counts the number of $1$s in $z$. All of these circuits can be simulated by $H(t)$ (1.1) in time*

$$T \le \mathrm{polylog}(N)N^{-1+\delta_{\mathrm{T}}}, \tag{5.9}$$

*with error*

$$\epsilon' = N\mathrm{polylog}(N)\epsilon, \tag{5.10}$$

*where $\epsilon$ is given in (3.2).*

When the Hamiltonian is 2-local with only $ZZ$ Ising terms, such circuits are known to be simulable in constant time [11]. Here we have shown that it can be even faster with $T \sim 1/N$, for $K$-local Hamiltonians.

*Proof.* All of the circuits can be built from the counting gate $\mathrm{Count} : |z\rangle |0 \cdots 0\rangle \to |z\rangle ||z|\rangle$ where $||z|\rangle$ is a bitstring on $m = \lfloor \log(n+1) \rfloor + 1$ ancilla qubits in set $S_{\mathrm{count}} = 1, \cdots, m$. Denote the $n$ qubits by set $S$. Following [10–12],

$$\mathrm{Count} = \mathrm{QFT}^\dagger \prod_{k=1}^m \mathrm{CZ}_{k,S}^{\theta_k} \mathrm{H}_k, \tag{5.11}$$

where $\mathrm{H}_k$ is Hadamard on $k$, $\mathrm{QFT}^\dagger$ is the inverse quantum Fourier transform (QFT) on $S_{\mathrm{count}}$, and $\theta_k = \pi/2^{k-1}$. To see (5.11), observe that the CZs map

$$|z\rangle \otimes_k \frac{|0\rangle + |1\rangle}{\sqrt 2} \to |z\rangle \otimes_k \frac{|0\rangle + \mathrm{e}^{\mathrm{i}\theta_k |z|} |1\rangle}{\sqrt 2}, \tag{5.12}$$

where the state on $S_{\mathrm{count}}$ is precisely the QFT states.

The counting gate (5.11) satisfies the condition in Corollary 5.2 with the number of gates (counting fan-out as one) $WD = \mathrm{polylog}(n)$, because the QFT is only on $\mathrm{O}(\log n)$ qubits. So it can be simulated by all-to-all Hamiltonian in time (5.9) and error $N_{\mathrm{d}}\mathrm{polylog}(n)\epsilon = N\mathrm{polylog}(N)\epsilon = \epsilon'$ because $n \le N_{\mathrm{d}} = \mathrm{O}(N)$. Note that we can only use $N - m = N - \mathrm{polylog}(N)$ ancillae to accelerate the gates, but this difference can be absorbed in the $\mathrm{polylog}(N)$ factors by e.g. $(N-m)^{-1+\delta_{\mathrm{T}}} \le 2N^{-1+\delta_{\mathrm{T}}}$.

Using Count and $\mathrm{Count}^\dagger$ for uncomputing, all gates in (5.8) are reduced to the logarithmic-size problem $||z|\rangle |y\rangle \to ||z|\rangle |y \otimes \widetilde{g}(|z|)\rangle$, which are implementable using poly $(m) = \mathrm{polylog}(n)$ two-qubit gates. As a result, these gates can also be simulated in time (5.9) and error (5.10). $\qquad\square$

### 5.3. Fast state preparation

Corollary 5.2 also implies that the GHZ state and W state can be prepared in time $\sim 1/N$.

**Example 5.4.** *Consider the same setting as Corollary 5.2. Let the data qubits be initially in the all-zero state $|0 \cdots 0\rangle$. The GHZ state $|\mathrm{GHZ}\rangle = \frac{1}{\sqrt{2}}(|0 \cdots 0\rangle + |1 \cdots 1\rangle)$ and W state $|\mathrm{W}\rangle = \frac{1}{\sqrt{N_d}}(|10 \cdots 0\rangle + |010 \cdots 0\rangle + \cdots + |0 \cdots 01\rangle)$ on the data qubits can be prepared in time* (5.9) *with error*

$$\left\| \mathcal{T} \mathrm{e}^{-\mathrm{i} \int_0^T dt H(t)} |0 \cdots 0\rangle \otimes |\mathbf{0}\rangle - |\psi_{\mathrm{tar}}\rangle \otimes |\mathbf{0}\rangle \right\| \leq \epsilon', \tag{5.13}$$

*where $\psi_{\mathrm{tar}} = \mathrm{GHZ}, \mathrm{W}$ and $\epsilon'$ is given in* (5.10).

*Proof.* The GHZ state is obvious from Corollary 5.2 because it can be prepared by a single fan-out gate.

For the W state, we first apply on-site rotations so that the data qubit state $|\theta \cdots \theta\rangle$ has $\Theta(1)$ overlap with $|\mathrm{W}\rangle$. This can be achieved by a uniform rotation $\prod_j \mathrm{e}^{\mathrm{i}\theta X_j}$ with angle $\theta = 1/\sqrt{N_d}$. We then apply a Grover-like algorithm to boost the overlap to nearly 1.

The Grover iteration requires reflection $I - 2 |\theta \cdots \theta\rangle \langle\theta \cdots \theta|$ about the initial state $|\theta \cdots \theta\rangle$, which is equivalent to reflection about the all-zero state up to on-site rotations. This reflection reduces to the Or gate and can be fast simulated in Example 5.3. Similarly, the reflection about the target state $|\mathrm{W}\rangle$ reduces to gate Exact[1], because the whole evolution remains in the DM of the data qubits where $|\mathrm{W}\rangle$ is the only state selected by Exact[1].

For the Grover iteration to boost to small error $\epsilon'/2$ without overshooting, we actually apply a fixed point version [13], where the reflections are replaced by angle-$\frac{\pi}{3}$ versions e.g.

$$I - \left(1 - \mathrm{e}^{\mathrm{i}\frac{\pi}{3}}\right) |\theta \cdots \theta\rangle \langle\theta \cdots \theta|. \tag{5.14}$$

These modified versions are also reducible to Or and Exact[1] by an extra constant amount of two-qubit gates, whose simulation has the same costs as Example 5.3.

Because the original overlap is $\Theta(1)$, it suffices to use $\mathrm{O}(\log \frac{1}{\epsilon'}) = \mathrm{polylog}(N)$ reflections to get a desired error $\epsilon'/2$ from the Grover algorithm [13]. Since each reflection can be simulated in time (5.9) with error $\epsilon'/[2 \times$ (number of iterations)], the whole protocol has the same asymptotic cost (5.9) with error $\epsilon'$. □

If one allows single-qubit measurements, then the W state can be prepared by replacing the Grover boost by a simple measurement on the ancilla of Exact[1] with $\Theta(1)$ success probability. If one further allows adaptive feedback based on the measurement outcome, then any Dicke state[3] can be prepared by the same cost (5.9) according to the protocol in [14], which calls the fast subroutine that simulates Exact[t]. The proof of these statements is direct and we leave it to the readers.

Another interesting class of circuits suitable for Corollary 5.2 is sequential quantum circuits [15–20], where the number of entangling gates per layer $W \ll N_d$ is subextensive. For example, the codewords of 2d toric code can be prepared by a sequential circuit in 2 spatial dimensions with $W = D = \Theta(\sqrt{N_d})$ [18, 21], i.e. linear system size. Corollary 5.2 then implies the toric code codewords can be prepared by all-to-all Hamiltonians in roughly constant time $T = \mathrm{O}(N_d^{\delta_T})$. However, this result is known from previous work using a different strategy: The codewords of the toric code, or any stabilizer code more generally, are preparable by a Clifford circuit, and any Clifford circuit can be exactly simulated in $T = \mathrm{O}(1)$ time [11]. It remains to see if there are interesting sequential circuits that go beyond Clifford, for which Corollary 5.2 implies a speed up over known results.

### 5.4. Tight Lieb-Robinson bounds in power-law interacting systems

The all-to-all Hamiltonian (1.1) is a special case of power-law interacting Hamiltonians [22], where instead of (1.2) we have

$$H_{\mathrm{int}}(t) = \sum_{S \in \Lambda} H_S(t), \qquad \sum_{S \supset \{i,j\}} \|H_S(t)\| \leq c_{\mathrm{pow}} \left(\mathrm{dist}(i,j)\right)^{-\gamma}, \tag{5.15}$$

where $H_S(t)$ acts nontrivially on set $S$ and $c_{\mathrm{pow}}$ is a constant. Here the qubits form a square lattice in $\mathrm{d}$ spatial dimensions with Hamming distance $\mathrm{dist}(i,j)$ between two qubits on the lattice. One can verify that (1.2) belongs to (5.15) for power-law exponent $\gamma = 0$. Note that (5.15) is special even comparing to the $\gamma = 0$ case of (5.15), because (5.15) allows $\Theta(1)$ coupling coefficients for certain $\geq$ 3-body interactions, which is forbidden in (1.2). It turns out the general form (5.15) obeys a fundamental speed limit:

---

[3] A Dicke state is an equal superposition of all bitstrings of weight $\mathsf{t}$, which includes the W state at $\mathsf{t} = 1$.

**Theorem 5.5** (A Lieb-Robinson bound for power-law interacting systems). *Consider any $H(t)$ (1.1) with $H_{\mathrm{int}}(t)$ given in (5.15) and $\gamma < \mathsf{d}$. For any two operators $\mathcal{O}_i, \mathcal{O}'_j$ supported on qubits $i, j$ with $\|\mathcal{O}_i\| = \|\mathcal{O}'_j\| = 1$,*

$$\frac{1}{2} \left\| [\mathcal{O}_i(T), \mathcal{O}'_j] \right\| \leq \exp\left( c_{\mathsf{d}} T N_{\mathrm{tot}}^{1-\frac{\gamma}{\mathsf{d}}} \right) - 1, \tag{5.16}$$

*where $\mathcal{O}_i(T)$ is $\mathcal{O}_i$ after Heisenberg evolution by $H(t)$ in time $T$, and $c_{\mathsf{d}}$ is a constant determined by $c_{\mathrm{pow}}$ and $\mathsf{d}$.*

Theorem 5.5 is proven for the special case of 2-local Hamiltonians in [23] (see also [24]). The proof for $K$-local ones is similar, which we present here for self-containedness:

*Proof.* We use Corollary 6 in [25]:

$$\frac{1}{2} \left\| [\mathcal{O}_i(T), \mathcal{O}'_j] \right\| \leq [\exp{(2Th)}]_{ij} \leq \sum_{\ell=1}^{\infty} \frac{(2T)^\ell}{\ell!} \left( h^\ell \right)_{ij}$$

$$\leq \sum_{\ell=1}^{\infty} \frac{(2T)^\ell}{\ell!} \|h\|^\ell = \mathrm{e}^{2T\|h\|} - 1, \tag{5.17}$$

where $h$ is a $N_{\mathrm{tot}} \times N_{\mathrm{tot}}$ matrix with elements $h_{ii} = 0$ and

$$h_{ij} = \sum_{S \supset \{i,j\}} \|H_S(t)\| \leq c_{\mathrm{pow}} \left( \mathrm{dist}(i,j) \right)^{-\gamma}, \tag{5.18}$$

from (5.15). $h$ basically quantifies the information transmission rates between all pairs of qubits.

(5.18) implies

$$\|h\| \leq \max_i \sum_j h_{ij} \leq \max_i \sum_j c_{\mathrm{pow}} \left( \mathrm{dist}(i,j) \right)^{-\gamma} \leq \frac{1}{2} c_{\mathsf{d}} N_{\mathrm{tot}}^{1-\frac{\gamma}{\mathsf{d}}}, \tag{5.19}$$

where constant $c_{\mathsf{d}}$ is determined by $c_{\mathrm{pow}}$ and $\mathsf{d}$. In (5.19), we have used the fact that the summation over $j$ is dominated by $\Theta(N_{\mathrm{tot}})$ $j$s with $\mathrm{dist}(i,j) = \Omega(N_{\mathrm{tot}}^{1/\mathsf{d}})$, due to $\gamma < \mathsf{d}$. This statement holds because there are at most $0.01 N_{\mathrm{tot}}$ $j$s that can be contained in the radius-$c'_{\mathsf{d}} N_{\mathrm{tot}}^{1/\mathsf{d}}$ ball centered at $i$ for a sufficiently small constant $c'_{\mathsf{d}}$. One can repeat the argument for the $0.01 N_{\mathrm{tot}}$ qubits so that most of them $0.99 \times 0.01 N_{\mathrm{tot}}$ are at distance $\geq c''_{\mathsf{d}} N_{\mathrm{tot}}^{1/\mathsf{d}}$ and thus subdominant in (5.19) comparing to the faraway qubits. Note that this statement holds regardless of the shape or side lengths of the whole lattice. Combining (5.17) with (5.19) yields (5.16). $\qquad \square$

(5.16) is known as a Lieb-Robinson bound [26], where the commutator needs to grow large in order for qubit $j$ at time $T$ feels an initial perturbation at $i$. (5.16) then implies that in the strongly long-range regime $\gamma < \mathsf{d}$, any Hamiltonian protocol cannot propagate information in timescales shorter than $T = \Theta(N_{\mathrm{tot}}^{-1+\frac{\gamma}{\mathsf{d}}})$. Our protocol Theorem 3.1 implies that this bound is tight up to an arbitrarily small difference $\delta$ in the exponent:

**Corollary 5.6** (Tightness of Lieb-Robinson bound). *Consider a rescaled version $H_\gamma(t) = N_{\mathrm{tot}}^{-\frac{\gamma}{\mathsf{d}}} H(N_{\mathrm{tot}}^{\frac{\gamma}{\mathsf{d}}} t)$ of the protocol $H(t)$ in Theorem 3.1. Putting the qubits on a $\mathsf{d}$-dimensional square lattice with side lengths $N_{\mathrm{tot}}^{\frac{1}{\mathsf{d}}}$, $H_\gamma(t)$ satisfies the power-law interaction constraint (5.15) for a constant $c_{\mathrm{pow}}$ determined by $\mathsf{d}, K$. Furthermore, its Heisenberg evolution yields*

$$\|[X_0(T), X_{-1}]\| \geq 1, \tag{5.20}$$

*for some*

$$T \leq \Theta(N_{\mathrm{tot}}^{-1+\frac{\gamma}{\mathsf{d}}+\delta_{\mathrm{T}}}), \tag{5.21}$$

*where*

$$\delta_{\mathrm{T}} = \frac{5}{\sqrt{K}-1}, \tag{5.22}$$

*is an arbitrarily small constant by choosing a large $K$.*

*Proof.* $H_\gamma(t)$ satisfies (5.15) because every 2-local interaction is set to the weakest coupling strength $N_{\text{tot}}^{-\frac{1}{d}}$ allowed in (5.15). The $\geq$ 3-local interactions are suppressed by extra powers of $N$ in (1.2) so are comparable to the 2-local ones when estimate the summation over $S$ in (5.15).

The rescaling enlarges the evolution time by $N_{\text{tot}}^{\frac{2}{d}}$ so (5.21) holds from (3.3). By setting (5.22), the error of the protocol $\epsilon = N^{-1/2} \ll 1$, so that

$$\|[X_0(T), X_{-1}]\| \geq \|\langle \mathbf{0}| [X_0(T), X_{-1}] |\mathbf{0}\rangle\|$$
$$= \left\|\langle \mathbf{0}| [\text{CZ}_{0,-1}^\dagger X_0 \text{CZ}_{0,-1}, X_{-1}] |\mathbf{0}\rangle\right\| + \text{O}(\epsilon) = \|\langle \mathbf{0}| [X_0 Z_{-1}, X_{-1}] |\mathbf{0}\rangle\| + \text{O}(\epsilon) = 2 + \text{O}(\epsilon) \geq 1, \quad (5.23)$$

which establishes (5.20). $\qquad \square$

### 5.5. Ancilla-assisted fast operator growth

The operator norm of commutators in (5.16) quantifies information propagation in the *worst case* scenario, i.e. whether there exists one initial state such that information propagates fast. In the context of quantum scrambling [24, 27–34], one is usually interested in the *average case*, quantified by the Frobenius norm instead:

$$\|\mathcal{O}\|_{\text{F}} := \sqrt{\frac{1}{\dim \mathcal{H}} \text{Tr}\left(\mathcal{O}^\dagger \mathcal{O}\right)}. \quad (5.24)$$

The Frobenius norm of commutators $[\mathcal{O}(T), \mathcal{O}']$ is known as out-of-time-ordered correlators (OTOC) at infinite temperature, which demonstrates a quantum version of chaos [35–37] when OTOC grows exponentially in time. OTOC is also closely related to the operator size size $[\mathcal{O}(T)]$ of $\mathcal{O}(T)$, which is the average weight when viewing $\mathcal{O}(T)$ as a "wavefunction" expanded over all Pauli strings [31, 32]. Therefore, the growth of OTOC can be interpreted as growth of operator size as the operator evolves in the operator space [38–40].

In particular, for 2-local all-to-all Hamiltonians with normalization (1.2), Corollary 1 in [41] proves that

$$T = \Omega(N_{\text{tot}}^{-1/2}) \quad (5.25)$$

in order for $\left\|[\mathcal{O}_i(T), \mathcal{O}_j']\right\|_{\text{F}} = \Theta(1)$. The scrambling time (5.25) is asymptotically longer than the worst-case signaling time $T \sim N_{\text{tot}}^{-1}$ for operator norms in (5.16). Such a separation has also been established in weak power-law interacting systems [42–44]. Our protocol in Corollary 5.2, however, shows that the separation no longer exists when one allows for $\Theta(N_{\text{tot}})$ ancilla qubits initialized and finalized in the all-zero state, which do not participate in the average trace in (5.24). This is because a $T = N_{\text{tot}}^{-1+\delta_{\text{T}}}$ time simulation of the unbounded fan-out gate grows the operator size maximally in the data-qubit Hilbert space

$$X_0 \rightarrow \prod_{\text{data qubit } j} X_j. \quad (5.26)$$

It would be interesting to see if this ancilla-assisted fast operator growth have implications in quantum scrambling contexts.

## 6. $\sqrt{N}$ SPEED-UP FOR ANY QUANTUM CIRCUIT

Corollary 5.1 shows that any quantum circuit can be simulated by an all-to-all Hamiltonian evolution in arbitrarily shorter time, given a space overhead proportional to the speed up. In Section 5.2 we have also shown that special families of circuits can be simulated $N$ times faster with constant space overhead.

In this section, we show that asymptotic speed up is actually possible for any circuit with constant space overhead, even for the minimal case of 2-local Hamiltonians[4], albeit with a milder speed up factor $\sqrt{N}$. We first prove this statement with a guarantee that the simulation is accurate for almost all input states.

---

[4] Note that the normalization (1.2) for 2-local Hamiltonians used in this section are much simpler: any pair of qubits interact with strength $\leq 1$, which does not depend on the system size.

**Theorem 6.1.** *Consider $N$ ancilla qubits and $2N$ data qubits with $N_{\mathrm{tot}} = 3N$. For any constants $0 < \delta_{\mathrm{T}} < 1/2$ and $\kappa > 0$, if $N$ is sufficiently large, then any quantum circuit $U_{\mathrm{cir}}$ of depth $D$ on the data qubits can be simulated by Hamiltonian* (1.1) *with $K = 2$ in time*

$$T \leq \widetilde{c}_{\delta_{\mathrm{T}},\kappa} N^{-\frac{1}{2}+\delta_{\mathrm{T}}} D, \tag{6.1}$$

*with error*

$$\mathbb{E}_\psi \left\| \left( \mathcal{T} \mathrm{e}^{-\mathrm{i} \int_0^T \mathrm{d}t\, H(t)} - U_{\mathrm{cir}} \right) |\psi\rangle \otimes |\mathbf{0}\rangle \right\|^2 \leq c_{\delta_{\mathrm{T}},\kappa} D^2 N^{-2\kappa}. \tag{6.2}$$

*Here $c_{\delta_{\mathrm{T}},\kappa}, \widetilde{c}_{\delta_{\mathrm{T}},\kappa}$ are constants determined by $\delta_{\mathrm{T}}, \kappa$, and $\mathbb{E}_\psi$ averages over any ensemble of states $\psi$ that form a 1-design $\mathbb{E}_\psi \left( |\psi\rangle \langle \psi| \right) = \frac{I}{\dim \mathcal{H}_{\mathrm{d}}} = 2^{-2N} I$.*

Although such a simulation fails for the worst-case inputs, it provides an "optimistic" version of the target unitary [45]. We then show that allowing classical randomness in the Hamiltonian protocol promotes the simulation to one that succeeds for *arbitrary* input with high probability:

**Corollary 6.2.** *Consider the same setting as Theorem 6.1. Any quantum circuit $U_{\mathrm{cir}}$ of depth $D$ on the data qubits can be simulated by Hamiltonian* (1.1) *with $K = 2$ in time* (6.1), *where the on-site fields $h_i^a(t)$ are allowed to be random. The simulation error*

$$\mathbb{E}_H \left\| \left( \mathcal{T} \mathrm{e}^{-\mathrm{i} \int_0^T \mathrm{d}t\, H(t)} - U_{\mathrm{cir}} \right) |\psi\rangle \otimes |\mathbf{0}\rangle \right\|^2 \leq c_{\delta_{\mathrm{T}},\kappa} D^2 N^{-2\kappa}, \tag{6.3}$$

*is small for any given input state $|\psi\rangle$, averaged over the randomness of the Hamiltonian $\mathbb{E}_H$.*

### 6.1. Optimistic unitaries

We will develop the ideas and present the proof in later subsections; here we first discuss some implications of the average-case result Theorem 6.1. (6.2) implies a $1/\mathrm{poly}(N)$ small Frobenius distance

$$
\begin{aligned}
\| \langle \mathbf{0} | U | \mathbf{0} \rangle - U_{\mathrm{cir}} \|_{\mathrm{F}}^2 &:= 2^{-2N} \mathrm{Tr} \left[ \left( \langle \mathbf{0} | U | \mathbf{0} \rangle - U_{\mathrm{cir}} \right)^\dagger \left( \langle \mathbf{0} | U | \mathbf{0} \rangle - U_{\mathrm{cir}} \right) \right] \\
&= 2^{-2N} \mathrm{Tr} \left[ \langle \mathbf{0} | \left( U - U_{\mathrm{cir}} \right)^\dagger | \mathbf{0} \rangle \langle \mathbf{0} | \left( U - U_{\mathrm{cir}} \right) | \mathbf{0} \rangle \right] \\
&\leq 2^{-2N} \mathrm{Tr} \left[ \langle \mathbf{0} | \left( U - U_{\mathrm{cir}} \right)^\dagger \left( U - U_{\mathrm{cir}} \right) | \mathbf{0} \rangle \right] \\
&\leq c_{\delta_{\mathrm{T}},\kappa} D^2 N^{-2\kappa},
\end{aligned}
\tag{6.4}
$$

where $U = \mathcal{T} \mathrm{e}^{-\mathrm{i} \int_0^T \mathrm{d}t\, H(t)}$, the trace is over data qubits, and the second-to-last line is equivalent to the left hand side of (6.2). $\langle \mathbf{0} | U | \mathbf{0} \rangle$ is therefore an optimistic quantum "unitary" of $U_{\mathrm{cir}}$ [45]. Note that the properties of optimistic unitaries in [45] generalize directly to the potentially non-unitary $\langle \mathbf{0} | U | \mathbf{0} \rangle$; it is close to unitary after all so for $\mathrm{poly}(N)$ calls of the simulation one can always focus on the $|\mathbf{0}\rangle$ output of the ancillae that happens with high probability.

Although the protocol in Theorem 6.1 could fail badly for some initial state $|\psi\rangle$, it could work well in practice where it is unlikely to encounter the rare worst-case events. An interesting example found in [45] is a version [46] of Shor's factoring algorithm [5], where quantum Fourier transform (QFT) subroutines are replaced by their optimistic versions while maintaining $\Omega(1)$ success probability for factoring. This yields a factoring circuit for $N$-bit numbers that has almost linear depth $\mathrm{O}(N^{1+\delta})$ using only $2N + \mathrm{O}(N/\log N)$ total qubits. The optimistic QFT used in [46] takes $\mathrm{O}(\log N)$ depth, and uses no ancilla comparing to previous non-optimistic (i.e. accurate for all inputs) constructions [47] using a large amount of ancillae. Here we can simulate the optimistic QFT in [46] by all-to-all Hamiltonians in $\sim 1/\sqrt{N}$ time and slightly more ancillae, and further reduce the time cost of Shor's algorithm:

**Example 6.3.** *For any constant $0 < \delta_{\mathrm{T}} < 1/2$, an $N$-bit number can be factorized with $\Omega(1)$ success probability by Hamiltonian evolution* (1.1) *with $K = 2$ in time*

$$T = \mathrm{O}\left( N^{\frac{1}{2}+\delta_{\mathrm{T}}} \right), \tag{6.5}$$

*using $3N + \mathrm{O}(N/\log N)$ total qubits.*

*Proof.* We apply the version of Shor's algorithm in Theorem 7 in [45] with the $\Theta(N)$ optimistic QFTs for modular exponentiation further replaced by our optimistic simulation in Theorem 6.1, which reduces the nearly linear depth to (6.5). Note that for (6.5) to hold we cannot implement the exact linear-depth QFT for the final period finding anymore. Instead, we use Theorem 4 in [45] which gives a non-optimistic QFT in $O(\log N)$ depth with only $N/\log N$ ancillae. Although this QFT is built from an optimistic one, we do not need to simulate it via our Hamiltonian protocol because its depth is already subdominant comparing to (6.5).

Our simulation adds only $N + O(N/\log N)$ extra ancilla qubits to the $2N + O(N/\log N)$ qubits. Using triangle inequality, our optimistic simulation of the optimistic circuit in [45] is still an optimistic QFT with $1/\text{poly}(N)$ small error, so our results follow from Theorem 7 in [45]. □

## 6.2. $\sqrt{N}$ speed-up for one gate

To demonstrate the idea of Theorem 6.1, we start from an alternative protocol to accelerate one two-qubit gate, which costs longer time $T \sim 1/\sqrt{N}$ than Theorem 3.1. We use this slower one because it can be generalized to simultaneously accelerates $\Theta(N)$ two-qubit gates, as we will see in later subsections. Intriguingly, this alternative protocol can be made exact even for 2-local Hamiltonians:

**Theorem 6.4.** *Consider $N$ ancilla qubits and 2 data qubits. There exists a 2-local all-to-all Hamiltonian (1.1) that produces CZ gate on the data qubits exactly*

$$\mathcal{T}e^{-i\int_0^T dt H(t)} |\psi\rangle \otimes |\mathbf{0}\rangle = CZ_{0,-1} |\psi\rangle \otimes |\mathbf{0}\rangle, \quad \forall \psi \tag{6.6}$$

*in time*

$$T \leq 3\sqrt{\pi/N}. \tag{6.7}$$

The idea of Theorem 6.4 is to apply the Mølmer-Sørensen (MS) scheme [48, 49] viewing the ancillae as one boson mode $B$ in Section 2. A version of the MS scheme uses time-dependent Hamiltonian

$$H_{\text{MS}}^{\text{b}}(t^{\text{b}}) = \begin{cases} \sqrt{2N}Z_0 \otimes \hat{p}, & t^{\text{b}} \in [0, T^{\text{b}}/4] \\ \sqrt{2N}Z_{-1} \otimes \hat{x}, & t^{\text{b}} \in [T^{\text{b}}/4, T^{\text{b}}/2] \\ -\sqrt{2N}Z_0 \otimes \hat{p}, & t^{\text{b}} \in [T^{\text{b}}/2, 3T^{\text{b}}/4] \\ -\sqrt{2N}Z_{-1} \otimes \hat{x}, & t^{\text{b}} \in [3T^{\text{b}}/4, T^{\text{b}}] \end{cases} \tag{6.8}$$

where we use $Z_0$ to control displacement in $\hat{x}$ direction like CD (3.11), or $Z_{-1}$ to control displacement in $\hat{p}$ direction.

For any bitstring $|z\rangle_{0,-1}$, the state of the boson returns to its initial state after the full protocol of duration $T^{\text{b}}$, because the displacement of the first (second) and third (fourth) stage cancel exactly. However, the state gains a nontrivial phase $e^{-i\Phi Z_0 Z_{-1}}$ depending on whether $Z_0 Z_1$ is even or odd. This is because bosonic displacement operators do not commute and gain a Berry phase

$$\Phi = 2(\sqrt{2N}T^{\text{b}}/4)^2 = \frac{1}{4}N(T^{\text{b}})^2, \tag{6.9}$$

proportional to the phase space area enclosed by trajectory $\alpha = (0,0) \rightarrow (\sqrt{2N}T^{\text{b}}/4, 0) \rightarrow (\sqrt{2N}T^{\text{b}}/4, \sqrt{2N}T^{\text{b}}/4) \rightarrow (0, \sqrt{2N}T^{\text{b}}/4) \rightarrow (0,0)$ for the $Z_0 = Z_{-1} = 1$ example.

Setting

$$T^{\text{b}} = \sqrt{\pi/N}, \tag{6.10}$$

yields $\Phi = \pi/4$ so that the protocol achieves $CZ_{0,-1}$ up to free single-qubit rotations. Following the reasoning behind Theorem 3.1, $H_{\text{MS}}^{\text{b}}(t^{\text{b}})$ can be simulated with polynomially small error by (1.1) for the $N+2$ qubits with $T = \Theta(T^{\text{b}}) = O(N^{-1/2})$.

To get an exact protocol with 2-local Hamiltonians, we do not simulate bosons from qubits. Instead, we generalize the MS scheme that works directly for qubits, as shown in Fig. 3(a).

*Proof of Theorem 6.4.* Similar to (6.8), our protocol contains 4 stages

$$H_{\text{MS}}(t) = \begin{cases} \frac{1}{2}Z_0 \otimes Y, & t \in [0, T/4] \\ \frac{1}{2}Z_{-1} \otimes X, & t \in [T/4, T/2] \\ -\varphi_Y Z_0 \otimes Y, & t \in [T/2, 3T/4] \\ -\varphi_X Z_{-1} \otimes X, & t \in [3T/4, T) \end{cases} \tag{6.11}$$

As we will choose

$$0 < \varphi_X, \varphi_Y < 1, \tag{6.12}$$

this Hamiltonian satisfies normalization (1.2) with $K = 2$.

Given bistring $z$ on the data qubits, the ancilla state just rotates in the DM and remains to be a coherent spin state $|\theta; \phi\rangle$ ($0 \leq \theta \leq \pi$) labeled by the solid angle. For example the initial state is $|\mathbf{0}\rangle = |0; \phi\rangle$, and is rotated to $|\frac{T}{4}; 0\rangle$ or $|\frac{T}{4}; \pi\rangle$ after the first stage of evolution. We choose $\varphi_X, \varphi_Y$ so that the ancillae return to the initial state $|\mathbf{0}\rangle$ after the full protocol, for all $z$. Due to symmetries $X \to -X$ and $Y \to -Y$ of the problem, it suffices to focus on $Z_0 = Z_{-1} = 1$ as the other $z$s would follow automatically.

Since the rotation angles $\propto T = \mathrm{O}(1/\sqrt{N}) \ll 1$, the evolution agrees with the boson version (6.8) in zeroth order, so $\varphi_X, \varphi_Y \approx 1/2$ does the job. More precisely, for the state $|\omega\rangle$ after the first two stages, there exists

$$\varphi_Y = \frac{1}{2} + \mathrm{O}(T), \tag{6.13}$$

that rotates the state to one with $\phi = \pi/2$, after the third stage. This state is of the form $|\frac{T}{2}\varphi_X; 0\rangle$ with

$$\varphi_X = \frac{1}{2} + \mathrm{O}(T), \tag{6.14}$$

so that the final stage with precisely the same $\varphi_X$ returns the state back to $|\mathbf{0}\rangle$.

Because the full evolution preserves $|\mathbf{0}\rangle$,

$$\mathcal{T}\mathrm{e}^{-\mathrm{i}\int_0^T \mathrm{d}t H_{\mathrm{MS}}(t)} = \mathrm{e}^{-\mathrm{i}\mathcal{O}_{\mathrm{MS}} \otimes Z}, \tag{6.15}$$

for some operator $\mathcal{O}_{\mathrm{MS}}$ on the data qubits that is a function of $Z_0, Z_{-1}$. From the $X \to -X$ and $Y \to -Y$ symmetries, and the fact that $Z_0, Z_{-1}$ only take two values, the only allowed form of $\mathcal{O}_{\mathrm{MS}}$ is

$$\mathcal{O}_{\mathrm{MS}} = \Phi_{\mathrm{MS}} Z_0 Z_{-1}, \tag{6.16}$$

for a scalar $\Phi_{\mathrm{MS}}$.

To find $\Phi_{\mathrm{MS}}$, we invoke Magnus expansion (see e.g. [50])

$$\mathcal{T}\mathrm{e}^{-\mathrm{i}\int_0^T \mathrm{d}t H(t)} = \exp\left[\Omega_1 + \Omega_2 + \mathrm{O}(h^3 T^3)\right], \quad \text{where}$$
$$\Omega_1 = -\mathrm{i}\int_0^T \mathrm{d}t\, H(t), \quad \Omega_2 = \frac{1}{2}\int_0^T \mathrm{d}t_1 \int_0^{t_1} \mathrm{d}t_2\, [H(t_1), H(t_2)]. \tag{6.17}$$

Here $h = \max_t \|H(t)\|$ and $\mathrm{O}(h^3 T^3)$ represents an operator whose operator norm $\leq c_{\mathrm{Mag}} h^3 T^3$ with a numerical constant $c_{\mathrm{Mag}}$. We do not plug in $H = H_{\mathrm{MS}}$ directly due to its large operator norm $\|H_{\mathrm{MS}}(t)\| = \Theta(N)$. Instead, we write $H_{\mathrm{MS}}(t) = \sum_i H_{\mathrm{MS},i}(t)$ as commuting terms, where each $H_{\mathrm{MS},i}(t)$ acts on the ancillae only at qubit $i$ and has $h = \mathrm{O}(1)$, so that

$$\mathcal{T}\mathrm{e}^{-\mathrm{i}\int_0^T \mathrm{d}t H_{\mathrm{MS},i}(t)} = \exp\left[\left(\frac{T}{4}\right)^2 [\tfrac{1}{2}Z_0 Y_i, \tfrac{1}{2}Z_{-1}X_i] + \mathrm{O}(T^3)\right] = \exp\left[-\mathrm{i}\frac{T^2}{32}Z_0 Z_{-1} Z_i + \mathrm{O}(T^3)\right], \tag{6.18}$$

where we only keep $\propto Z_i$ terms according to (6.15). Comparing to (6.16),

$$\Phi_{\mathrm{MS}} = \frac{T^2}{32} + \mathrm{O}(T^3) \geq \frac{T^2}{36}. \tag{6.19}$$

We set $\Phi_{\mathrm{MS}} = \frac{\pi}{4N}$ so that the protocol induces $\mathrm{CZ}_{0,-1}$ exactly (up to free single-qubit rotations). There exists $T$ satisfying (6.7) that achieves this due to (6.19). $\qquad \square$

As a remark, we require the initial state of the ancillae to be $|\mathbf{0}\rangle$, while the original MS scheme works for any initial bosonic state.

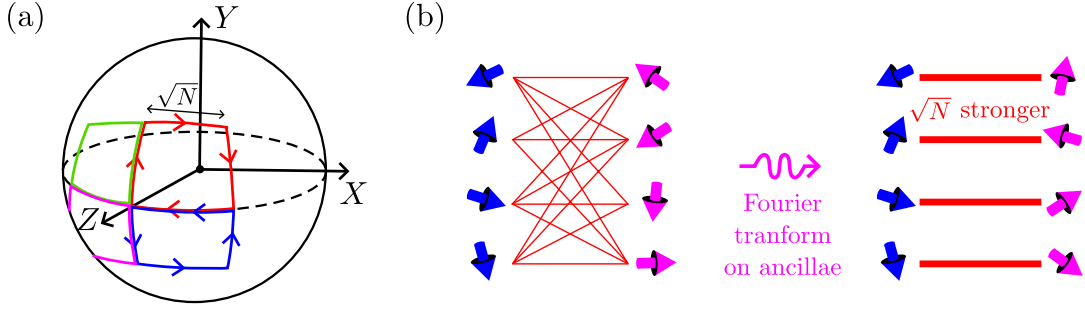

FIG. 3. (a) We apply the Mølmer-Sørensen scheme for qubits, which yields an exact $T = O(1/\sqrt{N})$ protocol of simulating one gate in Theorem 6.4. The four colored trajectories of the ancilla DM correspond to the four choices of $Z_0, Z_{-1} = \pm 1$, so that the system gains a geometric phase $\propto Z_0 Z_{-1}$ depending on the chirality of the trajectory. (b) For Theorem 6.1, we Fourier transform the ancilla qubits $X_i^+ \to \tilde{X}_k^+$, which focus the coupling strengths to make them stronger. Although the Fourier modes $\tilde{X}_k^+$ are not simple spins or bosons, they can be approximated so for almost all inputs. The idea is that each ancilla qubit $i$ starts from $|0\rangle$ and rarely visits its excited state $|1\rangle$ during the protocol, so each ancilla acts effectively as a boson working in its two lowest boson-number eigenstates.

### 6.3. Speeding up many gates simultaneously using ancilla bosons

To prove Theorem 6.1, we aim to simulate parallel CZ gates (1.5), where presumably the hardest case is $|S| = N$. Without loss of generality, we assume the $j$s involved are $-1, -2, \cdots, -N$.

We demonstrate the idea with a simpler setting, where each ancilla $i$ is a boson mode with operators $B_i, \hat{x}_i, \hat{p}_i$ instead of a qubit, and the coupling between a data qubit and an ancilla is $O(1)Z_j \otimes \hat{x}_i$ (or $\hat{p}_i$). Here one can pair each $j$ with one $i$ and apply the MS scheme to each pair in parallel, but this protocol has $\Omega(1)$ time because we lose the prefactor $\sqrt{N}$ in (6.8).

To regain the $\sqrt{N}$ prefactor, we transform the bosons into Fourier space[5]:

$$\mathcal{B}_k := \frac{1}{\sqrt{N}} \sum_{i=1}^{N} e^{i\frac{2\pi k}{N}i} B_i, \quad k = 1, \cdots, N \tag{6.20}$$

which are still boson annihilation operators $[\mathcal{B}_k, \mathcal{B}_{k'}^\dagger] = \delta_{kk'}$. The corresponding position/momentum operators are

$$\mathcal{X}_k = \frac{\mathcal{B}_k + \mathcal{B}_k^\dagger}{\sqrt{2}}, \quad \mathcal{P}_k = \frac{\mathcal{B}_k - \mathcal{B}_k^\dagger}{\sqrt{2}i}. \tag{6.21}$$

We now pair each Fourier boson $k$ with one data qubit $j = -k$, and engineer Hamiltonians like

$$H^{\mathrm{b}} = \sum_{k=1}^{N} \sqrt{2N} Z_{-k} \otimes \mathcal{X}_k = \sum_{k=1}^{N} Z_{-k} \otimes \sum_{i=1}^{N} e^{i\frac{2\pi k}{N}i} B_i + \mathrm{H.c.}, \tag{6.22}$$

which fulfills $O(1)$ coupling strength between each pair $(k, i)$. With the $\sqrt{N}$ prefactor as in (6.8), we can apply the MS scheme for each pair that simulates $N$ CZ gates in $O(1/\sqrt{N})$ time. The speed up comes from "focusing" $\Theta(N^2)$ couplings of strength $O(1)$ to $\Theta(N)$ couplings of strength $\Theta(\sqrt{N})$ by transforming to the Fourier modes. This $\sqrt{N}$ enhancement is optimal because the transformation keeps (# couplings) × (each coupling strength)$^2$ constant. This idea is illustrated in Fig. 3(b) for qubits that we will use.

In order to adapt this boson protocol to the original setting of qubit ancillae, let us first consider qudits with local Hilbert space dimension $d$, and truncate the above boson protocol to the qudit subspace. Here comes a crucial observation: This truncation error is small except for a tiny fraction of bitstrings $z$ of the data qubits. This can be understood from (6.22), where the maximal displacement for ancilla $i$ during the protocol is

$$\alpha_{\max}(i) = O(T_{\mathrm{MS}}) \sum_{k=1}^{N} e^{i\frac{2\pi k}{N}i} z_{-k} = O(N^{-\frac{1}{2} - \delta_{\mathrm{T}}}) \sum_{k=1}^{N} e^{i\frac{2\pi k}{N}i} z_{-k}, \tag{6.23}$$

---

[5] We avoid the name momentum space as $\hat{p}$ already means "momentum".

where $z_{-k}$ is the state on data qubit $-k$. Here we have assumed a slightly shorter MS duration $T_{\mathrm{MS}} = \mathrm{O}(N^{-\frac{1}{2} - \delta_{\mathrm{T}}})$ with arbitrarily small constant $\delta_{\mathrm{T}} > 0$, so that we need to repeat the MS scheme $\Theta(N^{2\delta_{\mathrm{T}}})$ times and get a mild overhead $N^{\delta_{\mathrm{T}}}$ in total time.

Although $\alpha_{\max}(i)$ can scale as $N^{\frac{1}{2} - \delta_{\mathrm{T}}} \gg 1$ in the worst case, $\alpha_{\max}(i) = \mathrm{O}(N^{-\delta_{\mathrm{T}}}) \ll 1$ for a random bitstring $z$ where each bit $z_{-k}$ is independently chosen as $\pm 1$ with $1/2$ probability. Moreover, the failure probability is exponentially small

$$\mathbb{P}\left[\left|\sum_{k=1}^{N} \mathrm{e}^{\mathrm{i}\frac{2\pi k}{N} i} z_{-k}\right| \geq \chi\right] \leq 2\mathbb{P}\left[\left|\mathrm{Re}\left(\sum_{k=1}^{N} \mathrm{e}^{\mathrm{i}\frac{2\pi k}{N} i} z_{-k}\right)\right| \geq \frac{\chi}{\sqrt{2}}\right] \leq 4\exp\left(-\frac{\chi^2}{4N}\right), \tag{6.24}$$

where we have used Hoeffding's inequality for the independent variables $-1 \leq \mathrm{Re}\left(\mathrm{e}^{\mathrm{i}\frac{2\pi k}{N} i} z_{-k}\right) \leq 1$. By a union bound on the $N$ choices of $i$

$$\mathbb{P}\left[\max_{i}\left|\sum_{k=1}^{N} \mathrm{e}^{\mathrm{i}\frac{2\pi k}{N} i} z_{-k}\right| \geq \chi\right] \leq 4N\exp\left(-\frac{\chi^2}{4N}\right), \tag{6.25}$$

with exponentially small failure probability we have

$$\alpha_{\max}(i) = \mathrm{O}(N^{-\delta_{\mathrm{T}}/2}) \ll 1, \tag{6.26}$$

for all $i = 1, \cdots, N$ from (6.23). As a result, truncating to the qudit subspace $\hat{n}_i \leq d-1$ incurs a polynomially small error $\mathrm{O}(N^{-\delta_{\mathrm{T}}(d-1)})$ for sufficiently large constant local dimension $d$, which one can verify from the coherent state expression (2.9).

The above strategy has an interesting analogy with spin waves in physics (see e.g. [6]): An interacting system of qudits/spins are usually beyond exact solvability, e.g. for the ground state of a Hamiltonian. However, when the local dimension $d$ is large, one can start from a classical ansatz state e.g. $|0\rangle$, and add quantum fluctuations to it suppressed in the small parameter $1/d$. The spin wave theory expresses each spin operator as bosons using the HP transformation, so that the dominant quantum fluctuation becomes a population of non-interacting bosons, usually solvable in Fourier (i.e. Bloch vector) basis. Although such a state with Fourier bosons populated is not contained in the original qudit Hilbert space, the spin wave theory is able to capture the essential physics of a large variety of systems, even for very small $d$ like qubits. Our reasoning above provides a rigorous bound on the truncation error of a "spin wave theory" in our context.

### 6.4. Proof of speed up for any quantum circuit

For the $d = 2$ qubit case, however, the truncation error $\mathrm{O}(N^{-\delta_{\mathrm{T}}})$ above is not good enough. Although one can use the DM of $d-1$ qubits as a qudit, it is not obvious how to simulate the boson protocol using qubit Hamiltonians even with $K$-locality. A direct simulation using HP transformation for each qudit would incur polynomially large poly $(N)$ coupling coefficients comparing to (1.2). Therefore, similar to Theorem 6.4, we analyze a qubit version of MS directly, rather than simulating some boson protocol.

*Proof of Theorem 6.1.* Using triangle inequality similar to (5.4),

$$\|(U - U_{\mathrm{cir}})|\psi\rangle \otimes |\mathbf{0}\rangle\| \leq \|(U_1 - U_{\mathrm{cir},1})|\psi\rangle \otimes |\mathbf{0}\rangle\| + \|(U_2 - U_{\mathrm{cir},2})U_{\mathrm{rot},1}U_{\mathrm{cir},1}|\psi\rangle \otimes |\mathbf{0}\rangle\|$$
$$+ \cdots + \|(U_D - U_{\mathrm{cir},D})U_{\mathrm{rot},D-1}U_{\mathrm{cir},D-1}\cdots U_{\mathrm{cir},1}|\psi\rangle \otimes |\mathbf{0}\rangle\|,$$
$$\mathbb{E}_\psi \|(U - U_{\mathrm{cir}})|\psi\rangle \otimes |\mathbf{0}\rangle\|^2 \leq D\mathbb{E}_\psi\left(\|(U_1 - U_{\mathrm{cir},1})|\psi\rangle \otimes |\mathbf{0}\rangle\|^2 + \|(U_2 - U_{\mathrm{cir},2})U_{\mathrm{cir},1}|\psi\rangle \otimes |\mathbf{0}\rangle\|^2 + \cdots\right)$$
$$\leq D^2 \max_m \mathbb{E}_\psi \|(U_m - U_{\mathrm{cir},m})|\psi\rangle \otimes |\mathbf{0}\rangle\|^2, \tag{6.27}$$

for $U = U_{\mathrm{rot},D}U_D\cdots U_2 U_{\mathrm{rot},1}U_1$, where we have used $U_{\mathrm{rot},1}U_{\mathrm{cir},1}|\psi\rangle$ for example also forms a 1-design. So it suffices to prove that each layer of CZ gates (1.5) can be simulated in time (6.1) and error (6.2) with $D = 1$ plugged in. We assume $|S| = N$ for now with $S = \{-1, -2, \cdots, -N\}$, and will see how to treat general $S$ in the end of the proof.

Similar to (6.22), consider $H^X = \sum_{i=1}^{N} H_i^X$ where

$$H_i^X = \frac{1}{2\sqrt{2}}\sqrt{N}\mathcal{Z}_i \otimes X_i^+ + \mathrm{H.c.} = \frac{1}{\sqrt{2}}\sum_{k=1}^{N} Z_{-k} \otimes \left[\cos\left(\frac{2\pi k}{N}i\right)X_i - \sin\left(\frac{2\pi k}{N}i\right)Y_i\right], \tag{6.28}$$

and

$$\mathcal{Z}_i = \frac{1}{\sqrt{N}} \sum_{k=1}^{N} \mathrm{e}^{\mathrm{i}\frac{2\pi k}{N}i} Z_{-k}, \tag{6.29}$$

so that $H_X$ satisfies (1.2) due to $|\cos\theta| + |\sin\theta| \le \sqrt{2}$. Define $H^Y = \sum_{i=1}^{N} H_i^Y$ similarly with

$$H_i^Y = \frac{\mathrm{i}}{2\sqrt{2}} \sqrt{N} \widetilde{\mathcal{Z}}_i \otimes X_i^+ + \mathrm{H.c.}, \tag{6.30}$$

and

$$\widetilde{\mathcal{Z}}_i := \frac{1}{\sqrt{N}} \sum_{k=1}^{N} \mathrm{e}^{\mathrm{i}\frac{2\pi k}{N}i} Z_{\tau(-k)}, \tag{6.31}$$

Note that in contrast to (6.21), $\mathcal{Z}_i$ is labeled by the ancilla index $i$ and represents a Fourier mode of the data qubits. From bound (6.25), we have

$$\max\left(|\mathcal{Z}_i|, |\widetilde{\mathcal{Z}}_i|\right) \le N^{\delta_\mathrm{T}/2}, \quad \forall i \tag{6.32}$$

when choosing each data qubit $Z_j = \pm 1$ in completely random, with exponentially small failure probability $\mathrm{e}^{-\Omega(N^{\delta_\mathrm{T}})}$. From now on we focus on these majority of data-qubit bitstrings for which (6.32) holds.

Mimicking the MS scheme, a primitive of our protocol is

$$V^{(1)}(t) = \mathrm{e}^{\mathrm{i}tH^X} \mathrm{e}^{\mathrm{i}tH^Y} \mathrm{e}^{-\mathrm{i}tH^X} \mathrm{e}^{-\mathrm{i}tH^Y} = \prod_i V_i^{(1)}(t), \tag{6.33}$$

which factorizes to commuting operators for each ancilla $i$:

$$V_i^{(1)}(t) = \exp\left[-t^2[H_i^X, H_i^Y] + E_i^{(1)}(t)\right] = \exp\left[\frac{\mathrm{i}}{2}\left(\mathcal{Z}_i^\dagger \widetilde{\mathcal{Z}}_i + \mathcal{Z}_i \widetilde{\mathcal{Z}}_i^\dagger\right) Nt^2 Z_i + E_i^{(1)}(t)\right], \tag{6.34}$$

using Magnus expansion (6.17). We will work with

$$t = \mathrm{O}(N^{-\frac{1}{2}-\delta_\mathrm{T}}), \tag{6.35}$$

so that $|t| \left\|H_i^X\right\|, |t| \left\|H_i^Y\right\| = \mathrm{O}(N^{-\delta_\mathrm{T}/2}) \ll 1$ due to (6.32). Therefore, the $\mathrm{O}(t^3)$ error term in (6.34) is bounded by

$$\left\|E_i^{(1)}(t)\right\| \le c^{(1)} N^{-3\delta_\mathrm{T}/2}, \tag{6.36}$$

for some numerical constant $c^{(1)} > 0$.

The error (6.36) is not small enough for our purpose. Nevertheless, the Suzuki method [51] recursively generates high-order product formulas for exponential of commutators [52–55]. Following Theorem 3 in [53], we define $V_i^{(p+1)}$ using $V_i^{(p)}$ by

$$V_i^{(p+1)}(t) = V_i^{(p)}(\gamma_p t) V_i^{(p)}(-\gamma_p t) \left(V_i^{(p)}(\beta_p t)\right)^{-1} \left(V_i^{(p)}(-\beta_p t)\right)^{-1} V_i^{(p)}(\gamma_p t) V_i^{(p)}(-\gamma_p t), \tag{6.37}$$

where

$$\beta_p := \sqrt{2r_p}, \quad \gamma_p := \sqrt{\frac{1}{4} + r_p}, \quad r_p := \frac{1}{4} \frac{2^{\frac{1}{p+1}}}{2 - 2^{\frac{1}{p+1}}}. \tag{6.38}$$

Iterating from the $p = 1$ case (6.34), we have [53]

$$V_i^{(p)}(t) = \exp\left[\frac{\mathrm{i}}{2}\left(\mathcal{Z}_i^\dagger \widetilde{\mathcal{Z}}_i + \mathcal{Z}_i \widetilde{\mathcal{Z}}_i^\dagger\right) Nt^2 Z_i + E_i^{(p)}(t)\right], \tag{6.39}$$

with error

$$\left\| E_i^{(p)}(t) \right\| \le c^{(p)} N^{-\delta_{\mathrm{T}}\left(p+\frac{1}{2}\right)}, \tag{6.40}$$

for a constant $c^{(p)} > 0$ determined by $p$, for any integer $p > 0$. Furthermore, when expressing $V_i^{(p)}(t)$ as evolutions using $\pm H_i^X, \pm H_i^Y$, costs time $\widetilde{c}^{(p)} t$ where $\widetilde{c}^{(p)} > 0$ is a constant determined by $p$.

Our whole protocol to simulate $U_{\mathrm{cir},m}$ is

$$U_m = \left( V^{(p)}(t) \right)^\nu, \quad \text{where} \quad V^{(p)}(t) = \prod_i V_i^{(p)}(t), \tag{6.41}$$

with

$$\nu = \left\lfloor N^{2\delta_{\mathrm{T}}} \right\rfloor, \quad t = \sqrt{\frac{\pi}{4N\nu}}, \tag{6.42}$$

so that the total time

$$T = \widetilde{c}^{(p)} t \nu \le \sqrt{\frac{\pi}{4}} \widetilde{c}^{(p)} N^{-\frac{1}{2}+\delta_{\mathrm{T}}}, \tag{6.43}$$

satisfies (6.1) with $D = 1$ and constant $\widetilde{c}_{\delta_{\mathrm{T}},\kappa} = \sqrt{\frac{\pi}{4}}\widetilde{c}^{(p)}$. We will determine $p$ by $\delta_{\mathrm{T}}, \kappa$ shortly in (6.47).

Using (6.39) and (6.40), $U = \mathrm{e}^{\mathrm{i}\Phi} + E$ with

$$\Phi \ket{\mathbf{0}} = \sum_i \left( \mathcal{Z}_i^\dagger \widetilde{\mathcal{Z}}_i + \mathcal{Z}_i \widetilde{\mathcal{Z}}_i^\dagger \right) = \frac{\pi}{4} \sum_{k=1}^N Z_{-k} Z_{\tau(-k)}, \tag{6.44}$$

where all contributions for $Z_{-k} Z_{\tau(-k')}$ with $k \ne k'$ cancel by the summation over $i$, and

$$\|E\| \le \nu N \max_i \left\| E_i^{(p)}(t) \right\| \le \nu N c^{(p)} N^{-\delta_{\mathrm{T}}\left(p+\frac{1}{2}\right)} \le c^{(p)} N^{1-\delta_{\mathrm{T}}\left(p-\frac{3}{2}\right)}. \tag{6.45}$$

Here (6.45) comes from Duhamel's inequality whose left hand side is bounded by $\int_0^T \mathrm{d}t' \, \|H_1(t') - H_2(t')\|$, and we set $H_1(t') - H_2(t')$ to be $\mathrm{i}t^{-1} \sum_i E_i^{(p)}$. Recall that these equations do not hold for all data qubit bitstrings $z$, but for the majority of them that ensures (6.32). We call these $z$s good. For any state 1-design in the whole Hilbert space, we have

$$\begin{aligned}
\mathbb{E}_\psi \left\| (U_m - U_{\mathrm{cir},m}) \ket{\psi} \otimes \ket{\mathbf{0}} \right\|^2 &= \mathbb{E}_z \left\| (U_m - U_{\mathrm{cir},m}) \ket{z} \otimes \ket{\mathbf{0}} \right\|^2 \\
&\le \mathbb{P}[z \text{ good}] \max_{z \text{ good}} \left\| (U_m - U_{\mathrm{cir},m}) \ket{z} \otimes \ket{\mathbf{0}} \right\|^2 + \mathbb{P}[z \text{ not good}] \cdot 2 \\
&\le \left( c^{(p)} \right)^2 N^{2-\delta_{\mathrm{T}}(2p-3)} + \mathrm{e}^{-\Omega(N^{\delta_{\mathrm{T}}})} \le 2 \left( c^{(p)} \right)^2 N^{2-\delta_{\mathrm{T}}(2p-3)},
\end{aligned} \tag{6.46}$$

which becomes (6.2) with $D = 1$ by setting

$$p = \left\lfloor \delta_{\mathrm{T}}^{-1}(\kappa+1) \right\rfloor + \frac{5}{2}, \tag{6.47}$$

and defining constant $c_{\delta_{\mathrm{T}},\kappa} = 2 \left( c^{(p)} \right)^2$. Here in (6.46) we have replaced any state 1-design with the particular 1-design of random bitstrings: $\mathbb{E}_\psi \|A\ket{\psi}\|^2 = \mathrm{Tr} \left[ A \mathbb{E}_\psi \left( \ket{\psi}\bra{\psi} \right) A^\dagger \right] = \mathrm{Tr} \left[ A \mathbb{E}_z \left( \ket{z}\bra{z} \right) A^\dagger \right] = \mathbb{E}_z \|A\ket{z}\|^2$, and used $\|(U_m - U_{\mathrm{cir},m}) \ket{z} \otimes \ket{\mathbf{0}}\| \le \|E\|$ because the $\mathrm{e}^{\mathrm{i}\Phi}$ contribution (6.44) agrees exactly with the target unitary, up to free single-qubit rotations.

We have proven (6.1) and (6.2) with $D = 1$ for the simulation of one layer (1.5), for the special case $|S| = N$. For $|S| < N$, we can just set the coefficients in (6.29) and (6.31) to zero for the qubits not acted by gates, and the analysis above still holds. We have thus proven the desired simulation for one layer (1.5); this finishes the whole proof for (6.1) and (6.2) because the time and error of different layers just add up (6.27). $\qquad\square$

### 6.5. Worst-case to average-case reduction

Here we prove Corollary 6.2 using standard methods of worst-case to average-case reductions [45, 56–58].

*Proof of Corollary 6.2.* For each CZ layer $U_{\mathrm{cir},m}$, Theorem 6.1 simulates it by $U_m$ that takes time (6.1) and has average error (6.2) with $D = 1$ plugged in.

Here, we follow [45] and simulate $U_{\mathrm{cir},m}$ by $\widetilde{V}^\dagger U_m V$ where $V = \prod_j V_j$ is on-site rotations on data qubits where each $V_j$ is randomly chosen from the four Paulis $I, X_j, Y_j, Z_j$ with equal $1/4$ probability, and

$$\widetilde{V}^\dagger := U_{\mathrm{cir},m} V^\dagger U_{\mathrm{cir},m}^\dagger, \tag{6.48}$$

is still a Pauli string, i.e. on-site rotations, because CZ gates in $U_{\mathrm{cir},m}$ are Clifford operations. It thus takes the same time to implement $\widetilde{V}^\dagger U_m V$ as $U_m$, from our assumption on fast on-site rotations. Since $V$ is a unitary 1-design, acting it on any fixed input $|\psi\rangle$ produces a state 1-design. This bounds the error

$$
\begin{aligned}
\mathbb{E}_V \left\| \left( \widetilde{V}^\dagger U_m V - U_{\mathrm{cir},m} \right) |\psi\rangle \otimes |\mathbf{0}\rangle \right\|^2 &= \mathbb{E}_V \left\| \widetilde{V}^\dagger \left( U_m - U_{\mathrm{cir},m} \right) V |\psi\rangle \otimes |\mathbf{0}\rangle \right\|^2 \\
&= \mathbb{E}_V \left\| \left( U_m - U_{\mathrm{cir},m} \right) V |\psi\rangle \otimes |\mathbf{0}\rangle \right\|^2 \\
&\leq c_{\delta_\mathrm{T}, \kappa} N^{-2\kappa},
\end{aligned}
\tag{6.49}
$$

where we have used $U_{\mathrm{cir},m} = \widetilde{V}^\dagger U_{\mathrm{cir},m} V$ in the first line, and that $V |\psi\rangle$ forms a state 1-design so we can apply (6.2) with $D = 1$.

Due to triangle inequalities similar to (6.27), the whole circuit $U_{\mathrm{cir}}$ can be simulated in time (6.1) with small error (6.3) for any initial $|\psi\rangle$.

$\square$

## 7. DISCUSSION ON THE FORM OF THE HAMILTONIAN

### 7.1. $K$-locality

Although $\sqrt{N}$-speed-up protocols in Section 6 apply to 2-local Hamiltonians, the $N$-speed-up protocol Theorem 3.1 needs $K$-local ones with a sufficiently large constant $K$ in order for the simulation error $\epsilon = \Theta(N^{-\kappa})$ to be as small as a desired polynomial. Here we argue that this $K$-locality requirement is not a severe problem.

First, there are experimental proposals to realize multibody interactions [59–63]. Interestingly, one naturally gains the $\propto N^{-k}$ normalization factor in (1.2) for $k$-body interactions.

Second, one can generate multibody interactions by Floquet engineering time-dependent 2-local Hamiltonians. For example, $\Omega_2$ in Magnus expansion (6.17) contains 3-body interactions if $H(t)$ is 2-local. This idea has been explored in [64], which proposes a 3-local time-independent Hamiltonian that generates a GHZ-like state in $\sim 1/N$ time. More generally, the order-$k$ terms in Magnus expansion involve nested commutators of $k$ $H(t)$s, integrated in $k$ time variables $0 < t_1, \cdots, t_k < T$, so can contain $(k+1)$-body interactions. Of course, this analysis is meaningful only if the Magnus expansion converges at least up to the desired order $k$. We estimate the strength of constant order $k$:

$$\|\Omega_k\| \sim T^k \|[\cdots [H(t_1), H(t_2)], \cdots, H(t_k)]\| \sim T^k N^{k+1}, \tag{7.1}$$

because by taking one commutator with $H(t)$, a constant-weight Pauli string maps to a superposition of $\sim N$ Pauli strings with O(1) coefficients, thus increasing the total norm by $\sim N$. (7.1) suggests that the Magnus expansion converges up to any constant order if

$$T \leq c'_{\mathrm{Mag}}/N, \tag{7.2}$$

for a sufficiently small constant $c'_{\mathrm{Mag}}$ [65]. The resulting effective Hamiltonian $\mathrm{i}T^{-1} \sum_k \Omega_k$ would match exactly with (1.2) in terms of the $N^{2-k}$ normalization prefactor. Although the Magnus expansion is expected to diverge at extremely large orders even assuming (7.2) because $T \|H(t)\| = \Theta(N) \gg 1$ (see (6.17)), the initial convergence at constant orders suggested by (7.2) may be good enough for our purpose. The idea comes from prethermalization theory [66–68], where an ultimate divergent Magnus expansion could faithfully describe the evolution up to a very long timescale. The above arguments support the possibility of obtaining effective $K$-local interactions from 2-local ones; however, it is unclear whether the desired form of $K$-local Hamiltonians like in Theorem 3.1 could be generated

by a 2-local protocol. We leave this as an interesting problem to explore in the future, where the Suzuki method used in the proof of Theorem 6.1 could be useful.

Finally, we need large enough $K$ for small error in our proof, but this does not rule out the possibility that simple 2-local protocols, which are not designed to simulate any $\geq$ 3-local interactions, could work well in practice. Indeed, [4] shows a simple 2-local protocol to fast generate the GHZ state, which has very small infidelity $< 10^{-3}$ for $\sim 10^3$ number of qubits. Crucially, the protocol utilizes spin squeezing [69, 70] generated by 2-local Hamiltonians, which is well established in numerics but escapes from rigorous treatments to the best of our knowledge. Although the infidelity in [4] seems to slowly increase when further increasing the system size, the protocol may be modified in easy ways to get better performance, e.g. using some of the ideas in this work. Such heuristic protocols could be more feasible in experiments, comparing to the rigorous but complicated protocols like Theorem 3.1.

### 7.2. Ising-type interactions and individually tunable couplings

(1.2) assumes an arbitrary form of interaction $\sum_{a_1 a_2 \cdots} X_{i_1}^{a_1} X_{i_2}^{a_2} \cdots$, while in many cases [11, 61, 71] it is more natural to restrict to Ising-type interactions where $H_{\text{int}}(t)$ only involves Pauli-$Z$s. Theoretically, this is not a problem because using fast on-site rotations, the system with only Ising-type interactions effectively feels Ising-type interactions along varying SU(2) directions that change rapidly in time, which could produce an effective Hamiltonian (1.2) with an arbitrary interaction form from first-order Magnus expansion. All high orders like $\Omega_2$ are suppressed by the arbitrarily fast on-site rotations. This idea appears in the literature [72], where two-axis twisting $H = XY + YX$ is simulated by the one-axis twisting Hamiltonian $H = Z^2$. In real experiments, however, it may take large effort to engineer interactions beyond Ising-type [73, 74].

Another feature of our protocols is that we need to carefully design the coupling matrix $J_{ij}$ (for 2-body interactions for example), instead of using e.g. uniform couplings. It is an interesting question to investigate whether such $J_{ij}$ could be realized efficiently in experiments (see recent developments [75–77]). Although $J_{ij}$s used in Theorem 6.1 seem to be quite featureless, those in Theorem 3.1 do couple the ancilla qubits in a uniform manner, which may simplify the problem. More generally, we leave the experimental feasibility of our protocols as future directions.

---



\end{document}